\documentclass[sigconf,10pt,nonacm]{acmart}

\usepackage{graphicx}
\usepackage{subcaption}         
\usepackage{booktabs}           
\usepackage{balance}            
\usepackage{xspace}
\usepackage{url}
\usepackage{mathtools}
\usepackage{enumitem}
\usepackage{algorithm}
\usepackage{algpseudocode}
\usepackage{microtype}
\algnewcommand\ServerInput{\item[\textbf{Server Input:}]}
\algnewcommand\ClientInput{\item[\textbf{Client Input:}]}

\begin{document}

\title{CARVY-FL: Client Anticlustering for Robust Voting in Provably Secure Federated Learning}

\author{Masaki Nakada}
\affiliation{%
  \institution{The University of Tokyo}
  \city{Tokyo}
  \country{Japan}}

\author{Honoka Anada}
\affiliation{%
  \institution{The University of Tokyo}
  \city{Tokyo}
  \country{Japan}}

\author{Tatsuya Kaneko}
\affiliation{%
  \institution{Institute of Science Tokyo}
  \city{Kanagawa}
  \country{Japan}}

\author{Hiroshi Nakamura}
\affiliation{%
  \institution{The University of Tokyo}
  \city{Tokyo}
  \country{Japan}}

\author{Shinya Takamaeda-Yamazaki}
\affiliation{%
  \institution{The University of Tokyo / RIKEN}
  \city{Tokyo}
  \country{Japan}}

\author{Hideki Takase}
\affiliation{%
  \institution{The University of Tokyo}
  \city{Tokyo}
  \country{Japan}}

\begin{abstract}
Federated learning (FL) enables IoT devices, acting as clients, to collaboratively train a model under the coordination of a central server without sharing raw data, thereby preserving privacy. However, FL remains vulnerable to malicious clients that can manipulate training and induce incorrect predictions. To quantify robustness against such attacks, prior work introduced Certified Accuracy (CA), where $\mathrm{CA}(m)$ denotes a lower bound on inference accuracy in the worst case with up to $m$ malicious clients. A promising approach to improving CA is voting-based FL, which forms multiple groups of clients, trains one model per group, and aggregates their predictions by plurality vote so that compromised groups can be outvoted by benign ones. However, under heterogeneous non-IID data, the existing random grouping approach, though efficient, can make CA highly sensitive to the particular grouping realization, resulting in inconsistent robustness guarantees.

To address this challenge, we propose CARVY-FL, a novel voting-based FL method designed to achieve high and stable CA under data heterogeneity and to improve empirical robustness against attacks. CARVY-FL introduces Anticlustering-based grouping to maximize distributional diversity within each group. It first infers each client's distribution type, i.e., the underlying type of local data distribution, from single-epoch model updates, then constructs diverse groups, trains a separate model for each group, and performs inference by plurality voting across the group models. By increasing distributional diversity within each group, CARVY-FL reduces the chance that particular groups become biased toward limited data distributions, thereby stabilizing group predictions and improving the reliability of plurality voting. Experiments on image classification tasks show that CARVY-FL achieves higher CA, tolerating nearly twice as many malicious clients as random grouping, while also improving empirical robustness under real attacks by about 11~\% relative to random grouping.
\end{abstract}


\maketitle

\section{Introduction}
\label{sec:intro}
Federated learning (FL)~\cite{mcmahan2017communication,bonawitz2019towards,kairouz2019advances} enables edge devices such as IoT devices~\cite{nguyen2021iotfl} to collaboratively train a model without sharing raw data, thereby preserving privacy by exchanging only model updates with a central server. However, FL is vulnerable to attacks by malicious clients, where a malicious client is either a fake client injected by an adversary or a legitimate client that has been compromised. A major threat in this setting is poisoning attacks, in which malicious clients manipulate either their local training data or the model updates sent to the server so as to bias the learned global model~\cite{bagdasaryan2020backdoor,analyzing_fl_adv_lens, fang2020bysattack,shejwalkar2021manipulating}. For example, in a backdoor-style poisoning attack such as BadNets~\cite{gu2017badnets, bagdasaryan2020backdoor}, an attacker lets several malicious clients participate in FL, and each of them trains on images stamped with a specific pattern called a trigger after relabeling those images as an attacker-chosen target class. The malicious clients then send the resulting model updates to the server. As a result, the trained model maintains high accuracy on clean inputs, while misclassifying triggered inputs as the attacker-intended target class.

To quantify robustness against such malicious clients in a provable manner, Certified Accuracy (CA) has been introduced~\cite{cao2022flcert}. $\mathrm{CA}(m)$ denotes a lower bound on inference accuracy in the worst case where up to $m$ malicious clients can induce arbitrary misclassifications. A promising approach to improving CA is voting-based FL, in which clients are assigned to one or more groups, FL is performed independently within each group to train one model per group, and the predictions of all group models are aggregated by plurality vote at inference time so that compromised groups can be outvoted by benign ones. Among grouping strategies, disjoint random grouping, typically realized via hash-based pseudorandom assignment, is attractive because it incurs low training cost; however, under non-IID settings it often yields highly variable and unstable CA~\cite{cao2022flcert}. One possible explanation for this instability is that hash-based assignment ignores client data distributions and therefore does not ensure sufficient distributional diversity within each group. This observation motivates client grouping methods that explicitly take client data distributions into account to achieve greater robustness.

We propose CARVY-FL (Client Anticlustering for Robust Voting in Provably Secure Federated Learning), which improves robustness by increasing the \emph{vote margin}, i.e., the gap between the number of votes for the correct class and that for the runner-up, through Anticlustering-based grouping. CARVY-FL estimates client distribution types from one-epoch model updates, forms groups containing clients from as many distribution types as possible, trains one model within each group, and performs inference by plurality voting across group models.

The contributions of this paper are as follows:
\begin{itemize}[leftmargin=*,nosep]
  \item We propose CARVY-FL, which replaces existing random grouping with Anticlustering-based grouping to achieve stronger attack robustness under class-disjoint non-IID settings.
  \item We introduce a privacy-aware method that estimates each client's distribution type from model updates, enabling Anticlustering without collecting raw data or explicit distribution information.
  \item We evaluate CARVY-FL using Certified Accuracy (CA)~\cite{cao2022flcert} and Attack Success Rate (ASR), and confirm that it improves robustness over both FLCert and a clustering-based grouping baseline.
\end{itemize}

The remainder of this paper is organized as follows. Section~\ref{sec:background} reviews background on FL, poisoning attacks, FLCert, and non-IID countermeasures. Section~\ref{sec:design} presents the design and algorithms of CARVY-FL. Section~\ref{chap:simulation} reports the experimental evaluation. Section~\ref{sec:conclusion} concludes the paper and discusses future directions.
\section{Background and Related Works}
\label{sec:background}

\subsection{Federated Learning}
Federated Learning (FL)~\cite{mcmahan2017communication,bonawitz2019towards,kairouz2019advances} is a framework in which numerous clients collaboratively train a model under the coordination of a central server while keeping their data on their own devices.

In each round $t$, the server selects participating clients $\mathcal{S}^t\subseteq\mathcal{N}$ from the client set $\mathcal{N}:=\{1,\dots,N\}$ and distributes the global model $w^t$ to each selected client. Each client $k\in\mathcal{S}^t$ performs local training on its own data, using the received model as initialization, and sends its updated model $w_k^{t+1}$, or the corresponding update, back to the server. In this paper, we collectively refer to the local models, their differences, gradient vectors, and similar quantities sent from clients to the server as \emph{updates}. The server aggregates the received updates to produce a new global model. In FedAvg~\cite{mcmahan2017communication}, the global model is updated by taking a weighted average of the local models, and is then redistributed to the clients. This process is repeated until convergence.

\subsection{Attacks by Malicious Clients}
Federated learning is vulnerable to attacks in which some participating clients are manipulated by an adversary. Such malicious clients can arise in two major forms. First, an attacker may inject fake clients into the system. Second, legitimate clients may be compromised by an attacker and made to send malicious updates. In either case, the attacker can intervene in the training process and intentionally distort the performance or behavior of the global model.

Attacks are broadly classified into untargeted attacks~\cite{fang2020bysattack,xie2020fallofempires,shejwalkar2021manipulating} and targeted attacks~\cite{analyzing_fl_adv_lens,shejwalkar2021manipulating,brauch2019alittle_enough,bagdasaryan2020backdoor}. Untargeted attacks aim to degrade overall predictive performance by manipulating updates to lower accuracy across the entire test dataset. Targeted attacks induce the attacker's intended misclassification only on specific inputs. Concretely, such attacks include those that force inputs with certain characteristics to be misclassified as a target label, as well as backdoor attacks~\cite{bagdasaryan2020backdoor} that trigger misclassification only when a special pattern (trigger) is embedded in the input, while maintaining high accuracy on ordinary inputs.

Untargeted attacks are relatively easy to detect because they manifest as accuracy drops on clean data, leaving room for defense mechanisms or retraining. In contrast, backdoor attacks such as BadNets~\cite{gu2017badnets} maintain normal accuracy on clean inputs, making them difficult to detect. Moreover, even a small number of malicious clients can induce a high misclassification rate~\cite{bagdasaryan2020backdoor}, so failure to detect them can have serious consequences. For this reason, targeted attacks, particularly backdoor attacks, are regarded as a more serious threat~\cite{ozdayi2021backdoorishard, Huang_2023_ICCV}.

In addition, model replacement attacks are known, in which the attacker adjusts malicious updates while taking the server's aggregation rule into account so as to strongly steer the aggregated global model toward the attacker's desired behavior~\cite{bagdasaryan2020backdoor}. By incorporating model replacement, even more powerful backdoor attacks can be realized.

Based on the above, this study considers a backdoor attack combined with model replacement by malicious clients as the primary threat model. This is because such an attack can achieve the attacker's objective without significantly degrading standard performance, making it difficult to detect and practically important as a threat to federated learning systems.

\subsection{Defenses against Malicious Clients}
Representative defense methods include robust aggregation in Byzantine-robust FL. For example, Krum~\cite{blanchard2017krum} selects the update closest to the others, trimmed mean~\cite{yin2018byzantine} removes extreme values in each dimension before averaging, and FLTrust~\cite{Cao2020FLTrust} performs weighted aggregation based on consistency with a server-held reference gradient.

However, many defense methods assume that malicious updates are statistically separable from benign ones, and can be evaded if the attacker adjusts updates to avoid appearing as outliers~\cite{fang2020bysattack}. To address this limitation, this study adopts Certified Accuracy (CA)~\cite{cao2022flcert} as a robustness metric, which provides a provable lower bound on inference accuracy in the presence of malicious clients.

\subsection{Certified Accuracy}
\label{sec:2_FLCert}

\subsubsection{Overview of Certified Accuracy and voting-based FL}

Certified Accuracy (CA)~\cite{cao2022flcert} is a metric that represents a lower bound on inference accuracy that is guaranteed even when at most $m$ malicious clients exist. That is, $\mathrm{CA}(m)$ gives a lower bound on the accuracy that is maintained no matter how the attacker behaves.

A promising framework for substantially improving CA is voting-based FL. In this framework, clients are assigned to one or more groups, and FL is performed independently within each group to train one group model per group. At inference time, the predictions of all group models are aggregated by plurality vote. Even if some groups are compromised by attackers, the final prediction may still be preserved as long as sufficiently many benign groups remain. CA is a metric for quantifying this property.

For client assignment, there are two types of hash-based pseudorandom grouping: disjoint grouping, in which each client is assigned to only one group, and overlapping grouping, in which assignment to multiple groups is allowed~\cite{cao2022flcert}. The latter can yield low CA for small $m$ because of its probabilistic behavior, and the computational cost of training many group models is also high~\cite{cao2022flcert}. Therefore, this study focuses on disjoint grouping.

\subsubsection{Disjoint hash-based pseudorandom grouping (random grouping)}

In the existing method, each client is assigned a unique identifier $\mathrm{ID}_k$, and a hash function taking that identifier as input pseudorandomly assigns each client to exactly one group~\cite{cao2022flcert}. Specifically, the group membership $\gamma(k)$ is defined as
\begin{eqnarray}
\gamma(k) := h(\mathrm{ID}_k)\in\{1,\dots,G\}
\end{eqnarray}
where $G$ is the total number of groups. The corresponding hash function is written as
\begin{eqnarray}
h:\{\mathrm{ID}_1,\dots,\mathrm{ID}_N\} &\to& \{1,\dots,G\}
\end{eqnarray}
Hereafter, we refer to this grouping method as \emph{random grouping}.

\subsubsection{Training and inference-time voting in voting-based FL}

In voting-based FL, after performing an arbitrary grouping, including the disjoint hash-based pseudorandom grouping described above, independent FL is carried out using only the clients belonging to each group $g\in\{1,\dots,G\}$, and a group model $w_g$ is obtained~\cite{cao2022flcert}.

At inference time, the input $x$ is fed to each group model $w_g$. Let $p_{g,c}(x)$ denote the logit score for class $c\in\mathcal{Y}$. The prediction of group $g$ is given by
\begin{eqnarray}
  \hat{y}_g(x) &:=& \arg\max_{c\in\mathcal{Y}} p_{g,c}(x)
\end{eqnarray}
The number of votes for class $c$ is defined as
\begin{eqnarray}
  v_c(x) &:=& \sum_{g=1}^{G}\mathbf{1}\!\left[\hat{y}_g(x)=c\right]
\end{eqnarray}
The final prediction by plurality vote, with ties broken by the smallest class index, is
\begin{eqnarray}
  \hat{y}_{\max}(x) &:=& \min \arg\max_{c\in\mathcal{Y}} v_c(x)
  \label{eq:plurality_tie_break}
\end{eqnarray}

Thus, even if the outputs of some groups are manipulated by attackers, the final prediction can be preserved as long as benign groups constitute the majority.

\subsubsection{Definition of vote margin and Certified Accuracy in disjoint grouping}

We now define the vote margin and Certified Accuracy in the setting of disjoint grouping~\cite{cao2022flcert}. Let $y$ be the true class, and define the class other than $y$ that receives the largest number of votes as
\begin{eqnarray}
  \hat{y}_{\neg y}(x) &:=& \min \arg\max_{c\in\mathcal{Y}\setminus\{y\}} v_c(x)
\end{eqnarray}
The vote margin, which represents how much the number of votes for the correct class exceeds the largest number of votes among all incorrect classes when all clients are benign, is defined as
\begin{eqnarray}
  \Delta(x)
  &:=& v_{y}(x) - v_{\hat{y}_{\neg y}(x)}(x)
     - \mathbf{1}\!\left[y>\hat{y}_{\neg y}(x)\right]
  \label{eq:vote_margin}
\end{eqnarray}
The indicator term reflects the tie-breaking rule in Eq.~\eqref{eq:plurality_tie_break}.

Next, suppose that at most $m$ clients are turned into malicious clients by an attacker. In disjoint grouping, each client belongs to at most one group, so a single malicious client can affect at most one group. The attacker is assumed to be able to fully control the output of any group to which the attacker belongs. The worst case occurs when attackers are placed in distinct groups that originally voted correctly: the correct class loses $m$ votes and a competing class gains $m$ votes, reducing the margin by $2m$. Thus the plurality-vote result is preserved when
\begin{eqnarray}
  \Delta(x) &\ge& 2m
\end{eqnarray}

Based on this, under the test data distribution $\mathcal{D}$, Certified Accuracy is defined as the lower bound on the accuracy of plurality vote when at most $m$ clients are malicious:
\begin{eqnarray}
  \mathrm{CA}(m)
  &=& \mathbb{P}_{(x,y)\sim\mathcal{D}}\!\left(
    \Delta(x) \ge 2m
  \right)
  \label{eq:certified_accuracy_final}
\end{eqnarray}
\cite{cao2022flcert}. In practice, $\Delta(x)$ is computed over a test dataset in an environment where only benign clients are present, and $\mathrm{CA}(m)$ is then evaluated from the resulting vote margins. Assuming that the data are sampled from the same distribution, this can be interpreted as a lower bound on the fraction of correct predictions that is guaranteed at deployment time even when up to $m$ clients are malicious.

\subsection{Data Heterogeneity and Countermeasures}
\label{sec:2_non_iid}
In FL, IID refers to the case where all clients' training distributions $\mathcal{D}_k^{\mathrm{train}}$ are identical; non-IID refers to the case where they differ. We call an extreme form where the class sets held by different subsets of clients are mutually disjoint \emph{class-disjoint}. In such a setting, multiple distinct data distributions coexist. Focusing on the class composition of each client's local data, we define a \emph{distribution type} as a subset of clients sharing the same class composition. For example, one subset of clients having only classes 0 and 1 while another has only classes 2 and 3 corresponds to a class-disjoint setting. A related concept is partially class-disjoint data (PCDD)~\cite{Li2022nonIIDsiloPCDD,Fan2023PCDD}, though this study considers the more challenging fully class-disjoint setting.

Under non-IID environments, methods such as FedAvg that average local updates are known to suffer from client drift---systematic bias caused by the discrepancy between each client's local objective function $F_k$ and the global objective function $F$---resulting in delayed convergence, degraded accuracy, and oscillation across rounds~\cite{karimireddy2020scaffold}. This phenomenon becomes particularly severe in class-disjoint settings, where gradient bias is large.

One countermeasure is to group clients with similar distributions and perform FL within each group, as in Clustered FL~\cite{sattler2021clustered}. In this paper, we refer to this as clustering-based grouping. When clients with similar distributions are trained together, their gradient directions tend to be more consistent, which is expected to mitigate client drift.

Another representative countermeasure is SCAFFOLD~\cite{karimireddy2020scaffold}, which addresses performance degradation under non-IID directly through the learning algorithm itself. SCAFFOLD suppresses client drift by correcting the bias of local updates using control variates, introducing a server-side control variate $c$ and client-side control variates $c_k$.

\subsection{Challenges When Data Diversity and Hard-to-Detect Attacks Coexist}
\label{sec:2_joint_setting}
In this section, we summarize the limitations of existing methods in settings where client data distributions differ substantially and hard-to-detect attackers are present.

Certified voting-based defenses that do not rely on detection~\cite{cao2022flcert} use random grouping based on hash-based pseudorandom assignment. There are also studies that apply this mechanism to federated reinforcement learning in non-IID environments~\cite{fang2025provablyrobustfederatedreinforcement}. However, because this grouping does not take client data distributions into account, in class-disjoint settings CA becomes highly sensitive to the grouping realization and exhibits large variance~\cite{cao2022flcert}.

Replacing random grouping with clustering-based grouping stabilizes per-group learning but can harm plurality-vote accuracy under class-disjoint settings. Even if each group model achieves high accuracy on its own distribution, it may fail to learn decision boundaries spanning different distribution types. For example, if one group has only classes 0--1 and another only classes 2--3, an image of class~2 that is visually close to class~1 would be misclassified by the first group and correctly classified by the second, resulting in a tie rather than a correct prediction. In other words, no model has learned the decision boundary between classes 1 and 2, so this does not necessarily lead to improved plurality-vote accuracy on a uniform test distribution.

Furthermore, non-IID countermeasures such as SCAFFOLD stabilize within-group learning but do not address the group design problem itself, namely, which distribution types should be mixed within each group. Improving only the learning algorithm is insufficient to systematically enlarge the vote margin and achieve high and stable CA.

From the above, simply using existing methods alone or combining them naively is insufficient to simultaneously achieve certified robustness against hard-to-detect attacks and stable CA under high data diversity such as class-disjoint settings. To bridge this gap, a new grouping strategy is needed that can stably enlarge the vote margin while preserving the certified framework based on plurality vote.

\section{CARVY-FL Design}
\label{sec:design}

In this section, we propose CARVY-FL (Client Anticlustering for Robust Voting in Provably Secure Federated Learning), a robust federated learning method that withstands stealthy malicious clients even in environments where data distributions vary significantly across clients, particularly in class-disjoint settings. CARVY-FL retains the inference-time plurality voting and certified accuracy framework of voting-based FL \cite{cao2022flcert}, while replacing the random client grouping strategy with a distribution-aware grouping scheme, thereby boosting the vote margin under data heterogeneity.

\subsection{Problem Setting}
\label{sec:3_scenario}
We consider a classification task under federated learning, with the goal of maintaining inference accuracy even in the presence of malicious clients.
The server is assumed to be honest and to operate faithfully according to the protocol for model distribution and aggregation. The server cannot observe the raw training data of any client, nor can it directly obtain the per-client class distribution.
Benign clients participate in training following the protocol. For simplicity, we consider the case where all clients participate in every round ($\mathcal{S}=\mathcal{N}$). The data distributions of benign clients are assumed to be class-disjoint, meaning that the sets of classes held by different distribution types are mutually exclusive. Finally, malicious clients---adversaries lurking among the participants---may collude when multiple such clients are present. We consider scenarios ranging from a single malicious client to multiple malicious clients. Attacks are carried out by manipulating model updates.
The empirical impact of adversarial group placement is partially addressed in our ASR evaluation (Section~\ref{sec_4:evaluation}).

\subsection{Overview of the Proposed Method}
\label{sec:3_proposed_method}
\begin{figure*}[t]
  \centering
  \includegraphics[width=1\linewidth]{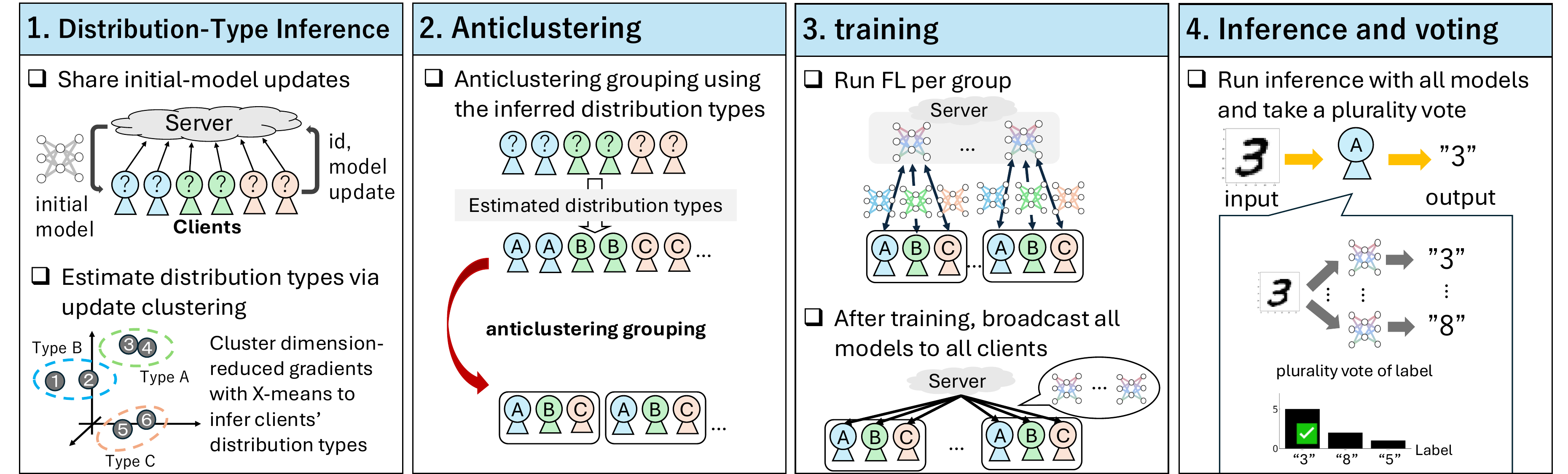}
  \caption{The four stages of the proposed method CARVY-FL: distribution-type inference, Anticlustering-based grouping, per-group training, and inference-time plurality voting.}
  \Description{Diagram illustrating the four stages of CARVY-FL: Step 1 collects one-epoch updates and clusters them, Step 2 applies Anticlustering to form diverse groups, Step 3 trains models within each group, and Step 4 aggregates predictions via plurality voting.}
  \label{fig:CARVY_FL_overview}
\end{figure*}

CARVY-FL builds upon voting-based FL's plurality-vote inference by replacing the group construction step with Anticlustering, which accounts for client data distributions. As illustrated in Figure~\ref{fig:CARVY_FL_overview}, CARVY-FL consists of four stages. First, the server collects one-epoch update differences from all clients, clusters them, and obtains estimated distribution-type clusters (Step 1). Next, Anticlustering is performed to construct groups such that each group contains clients with diverse distribution types (Step 2). Then, federated learning is executed independently within each group to obtain group models (Step 3). To mitigate client drift that may arise from intra-group distribution diversity, training employs SCAFFOLD \cite{karimireddy2020scaffold}. Finally, at inference time, all group models produce predictions and the final output is determined by plurality vote (Step 4).

\subsection{Distribution-Type Inference}
\label{sec:3_dist_inference}
In this section, we describe Step~1, which estimates the distribution type to which each client belongs based on model updates. This method exploits the property that, in class-disjoint environments, clients with different distribution types tend to produce systematically different update differences after one epoch of training. Since only update differences are used, the estimation is performed without collecting raw training data.

As initialization, the server holds the initial model $w^{0}$, the target dimensionality $d$ for principal component analysis (PCA), and a random seed $s$. Each client $k$ holds its local training data $\mathcal{D}_k^{\mathrm{train}}$.
Each client trains for one epoch starting from $w^{0}$ and sends $\Delta w_k=\mathrm{vec}(w_k^{(1)}-w^{0})$ to the server. The server stacks $\{\Delta w_k\}$ into a matrix, compresses it to $d$ dimensions via PCA, and applies X-means \cite{Pelleg2000Xmeans} to the resulting representations to automatically estimate the number of clusters. The output is the set of estimated distribution-type clusters $\mathcal{H}=\{h_1,\dots,h_L\}$ (Algorithm~\ref{alg:dist_xmeans}).

\begin{algorithm}[t]
\caption{Distribution-Type Inference via One-Epoch Training + X-means}
\label{alg:dist_xmeans}
\begin{algorithmic}[1]
\ServerInput clients $\mathcal{N}$, initial model $w^{0}$, PCA compressed dimension $d$, seed $s$
\ClientInput local train datasets $\{\mathcal{D}_k^{\mathrm{train}}\}_{k\in\mathcal{N}}$
\Ensure inferred clusters $\mathcal{H}=\{h_1,\dots,h_L\}$ (distribution-type clusters)

\For{\textbf{each} client $k\in\mathcal{N}$}
  \State Distribute and load $w^{0}$ into client $k$
  \State Locally train $w^{0}$ on $\mathcal{D}_k^{\mathrm{train}}$ for one epoch to obtain $w_k^{(1)}$
  \State $\Delta w_k \gets \mathrm{vec}(w_k^{(1)} - w^{0}) \in \mathbb{R}^{P}$
  \State Send $\Delta w_k$ to the server
\EndFor

\State Construct $F \gets [\Delta w_1^\top;\Delta w_2^\top;\dots;\Delta w_N^\top]\in\mathbb{R}^{N\times P}$
\State $Z \gets \mathrm{PCA}(F, d)$ \Comment{$Z\in\mathbb{R}^{N\times d}$}
\State Run X-means on $Z$ with seed $s$ to get number of clusters $L$ and clusters $\mathcal{H}=\{h_1,\dots,h_L\}$
\end{algorithmic}
\end{algorithm}

\subsection{Anticlustering-Based Grouping}
\label{sec:3_anticlustering}
In this section, we describe Step~2, which constructs the group set $\mathcal{G}$ using the estimated distribution-type clusters $\mathcal{H}$. The objective is to provide a grouping strategy that increases the vote margin at inference-time plurality voting in class-disjoint environments, thereby improving certified accuracy.

\subsubsection{Grouping Strategy for Increasing Vote Margin}
\label{subsec:3_1_group_vs_CA}
Under the class-disjoint assumption, for any class $y$, increasing the number of groups $G_y$ that contain a client holding class-$y$ data increases the vote count for the correct class $y$, improving the first term $v_{y}(x)$ of the vote margin in Equation~\eqref{eq:vote_margin} and thus increasing the vote margin. Conversely, having fewer groups that lack class $y$ reduces $v_{\hat{y}_{\neg y}(x)}(x)$, which further improves the vote margin. Summarizing this discussion, the grouping strategy expected to maximize the vote margin is one where, for any class $y$, the number of groups containing that class is large and the number of groups lacking it is small.
We note that the above argument is an intuitive hypothesis rather than a formal guarantee: it assumes that a group model trained with class-$y$ data will predict class $y$ correctly, which depends on factors such as training quality and inter-class similarity. Our experimental evaluation in Section~\ref{chap:simulation} provides empirical support for this hypothesis.
Reinterpreting this principle from the perspective of grouping clients in a class-disjoint setting, the ideal grouping assigns exactly one client from each distribution type to each group, so that every group contains as many distribution types as possible. In this paper, we refer to this construction as Anticlustering grouping and adopt it in CARVY-FL. See Figure~\ref{fig:method_comparision_all} for an illustration of Anticlustering grouping.

\subsubsection{Anticlustering Procedure}
\begin{figure}[t]
  \centering
  \includegraphics[width=\linewidth]{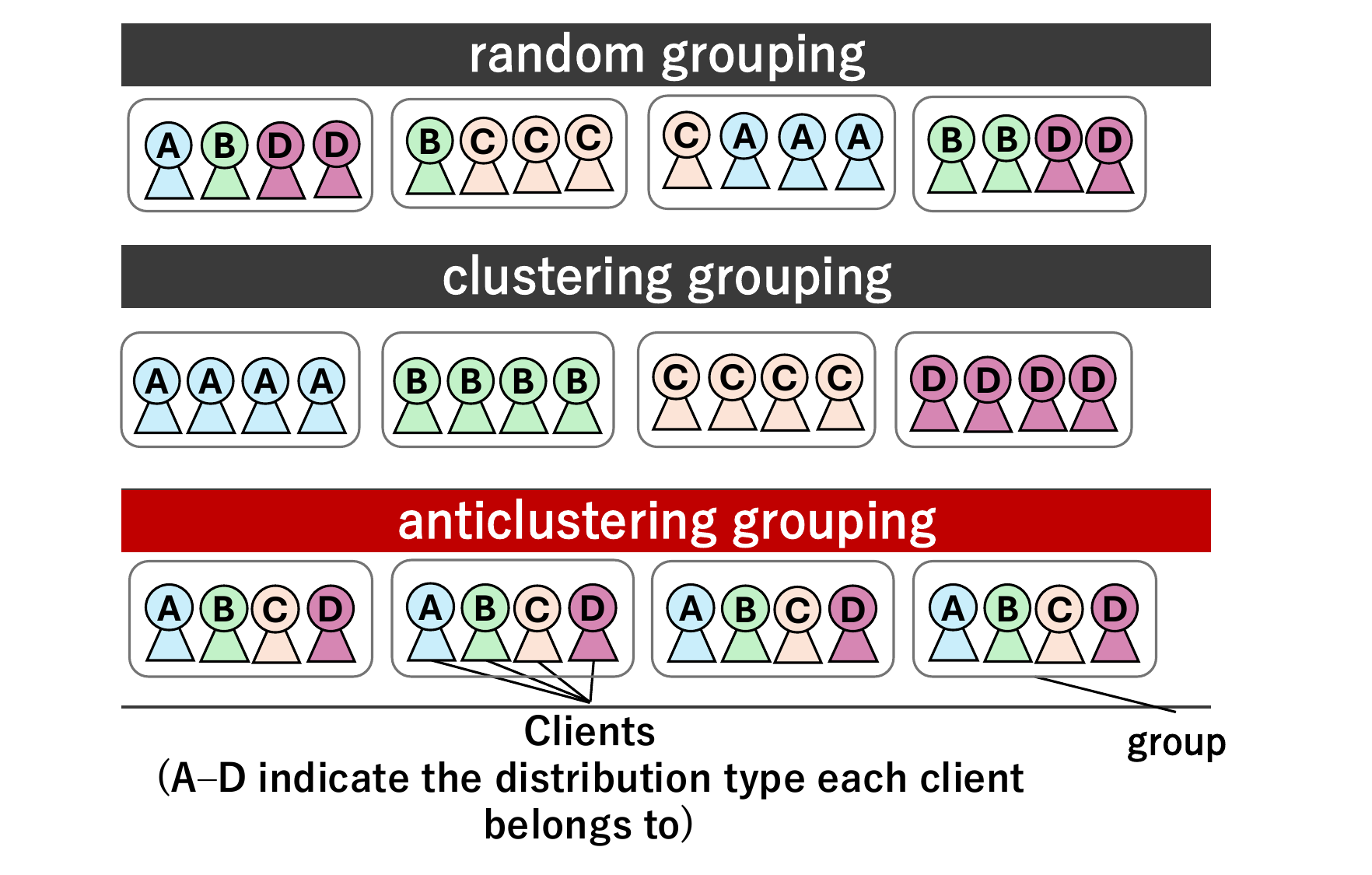}
  \caption{Comparison of grouping strategies: random grouping, clustering, and Anticlustering. (Note: Random grouping assigns clients based on hash values, so in practice the number of clients per group varies.)}
  \Description{Diagram comparing three grouping strategies: random grouping produces uneven and mixed groups, clustering isolates distribution types into separate groups, and Anticlustering distributes all distribution types evenly across groups.}
  \label{fig:method_comparision_all}
\end{figure}

Based on the above discussion, we construct the Anticlustering scheme that is expected to improve adversarial robustness (Algorithm~\ref{alg:carvy_anticlustering}). The input consists of the estimated clusters $\mathcal{H}=\{h_1,\dots,h_L\}$. First, the ordering within each estimated distribution-type cluster is randomly permuted (lines~1--3). Then, groups are formed by repeatedly selecting one client from each non-empty cluster and assigning them to the same group, thereby constructing the group set $\mathcal{G}$ (lines~4--14). This procedure ensures that each group contains as diverse a set of distribution types as possible while avoiding placing multiple clients from the same cluster into the same group. The number of groups is automatically determined by $\max_\ell |h_\ell|$. Figure~\ref{fig:method_comparision_all} illustrates examples of random grouping, clustering, and Anticlustering. The advantage of Anticlustering is that it is expected to boost the vote margin and certified accuracy of the plurality vote. On the other hand, because the grouping relies on client-derived update information, an adversary may manipulate its updates to influence the estimated distribution-type cluster to which it is assigned; this is a disadvantage compared to random grouping. This drawback is shared with clustering-based grouping.

\begin{algorithm}[t]
\caption{\textsc{BuildAnticlusterGroups} (Step 2): Anticlustering from inferred clusters}
\label{alg:carvy_anticlustering}
\begin{algorithmic}[1]
  \Statex \textbf{Input:} inferred clusters $\mathcal{H}=\{h_1,\dots,h_L\}$ where $h_\ell \subseteq \mathcal{N}$
  \Statex \textbf{Output:} FL groups $\mathcal{G}=\{g_1,\dots,g_G\}$, a partition of $\mathcal{N}$

  \For{$\ell=1$ \textbf{to} $L$}
    \State randomly permute $h_\ell$
  \EndFor

  \State $\mathcal{G}\gets \emptyset$
  \While{$\exists \ell \ \text{s.t.}\ h_\ell \neq \emptyset$}
    \State $g \gets \emptyset$
    \For{$\ell=1$ \textbf{to} $L$}
      \If{$h_\ell \neq \emptyset$}
        \State select and remove one client $u$ from $h_\ell$
        \State $g \gets g \cup \{u\}$
      \EndIf
    \EndFor
    \State $\mathcal{G} \gets \mathcal{G} \cup \{g\}$
  \EndWhile
\end{algorithmic}
\end{algorithm}

\subsection{Per-Group Training}
\label{sec:3_train}
In this section, we describe Step~3, in which federated learning is performed independently within each group $g\in\mathcal{G}$ obtained via Anticlustering to produce group models $\{w_g\}$ (Algorithm~\ref{alg:train_per_group}).

To mitigate client drift that may arise from intra-group distribution diversity, training within each group employs SCAFFOLD \cite{karimireddy2020scaffold}.
First, the initial global model and the SCAFFOLD server control variable for each group are initialized (lines~1--3).
Client control variables are also initialized (lines~4--6). Training then proceeds independently for each group with early stopping (lines~10--12). Groups can be trained in parallel. The training algorithm uses SCAFFOLD (line~13, Algorithm~\ref{alg:scaffold_round_slim}). For convergence detection, the mean validation accuracy $A_g^r$ across clients in the group is used; training terminates early if the improvement in $A_g^r$ remains below a threshold for a prescribed number of consecutive rounds (lines~14--18).

\begin{algorithm}[t]
\caption{Group-wise training with SCAFFOLD (outer loop)}
\label{alg:train_per_group}
\begin{algorithmic}[1]
  \ServerInput initial weights $w^{0}$; max rounds $R$; local steps $K$; local step-size $\eta_\ell$; global step-size $\eta_{\mathrm{global}}$; early-stop tolerance $\varepsilon$, patience $P$
  \ClientInput each client $i\in\mathcal{N}$ has:
    its FL group $g \in \mathcal{G}$ such that $i\in g$,
    and local datasets $(\mathcal{D}_i^{\mathrm{train}}, \mathcal{D}_i^{\mathrm{val}})$
  \Ensure trained group models $\{w_g\}_{g \in \mathcal{G}}$

  \For{$g$ \textbf{in} $\mathcal{G}$}
    \State initialize group model and server control: $w_g \gets w^{0}$;\ $c_g \gets \mathbf{0}$
  \EndFor
  \For{\textbf{each} client $i\in\mathcal{N}$}
    \State initialize client control: $c_i \gets \mathbf{0}$
  \EndFor

  \For{$g$ \textbf{in} $\mathcal{G}$} \Comment{groups can run in parallel}
    \State $best \gets -\infty$;\ $no\_improve \gets 0$
    \For{$r=0$ \textbf{to} $R-1$}
      \If{$no\_improve \ge P$}
        \State \textbf{break}
      \EndIf
      \State $(w_g, c_g, \{c_i\}_{i\in g}, A_g^r) \gets$
      \Statex \hspace{1.7em}\textsc{GroupSCAFFOLDRound}$(g, w_g, c_g, \{c_i\}_{i\in g}; K,\eta_\ell,\eta_{\mathrm{global}})$
      \Comment{Algorithm~\ref{alg:scaffold_round_slim}}
      \If{$A_g^r > best + \varepsilon$}
        \State $best \gets A_g^r$;\ $no\_improve \gets 0$
      \Else
        \State $no\_improve \gets no\_improve + 1$
      \EndIf
    \EndFor
  \EndFor
\end{algorithmic}
\end{algorithm}

\begin{algorithm}[t]
\caption{\textsc{GroupSCAFFOLDRound}: one SCAFFOLD round within a group}
\label{alg:scaffold_round_slim}
\begin{algorithmic}[1]
  \ServerInput group $g$; model $w$; server control $c$; client controls $\{c_i\}_{i\in g}$; $K,\eta_\ell,\eta_{\mathrm{global}}$
  \ClientInput each client $i\in g$ has datasets $(\mathcal{D}_i^{\mathrm{train}}, \mathcal{D}_i^{\mathrm{val}})$
  \Ensure updated $(w^+, c^+, \{c_i^+\}_{i\in g})$ and validation accuracy $A$

  \State \textbf{Server broadcasts} $(w,c)$ to all clients $i\in g$
  \For{\textbf{each} client $i\in g$ \textbf{in parallel}}
    \State $(w_i^+, c_i^+, a_i) \gets \textsc{SCAFFOLDClientUpdate}(i, w, c, c_i; K,\eta_\ell)$
    \Comment{standard SCAFFOLD client update \cite{karimireddy2020scaffold}}
  \EndFor

  \State \textbf{Server aggregates} $\{w_i^+\}_{i\in g}$ to update $w^+$ (standard SCAFFOLD \cite{karimireddy2020scaffold})
  \State \textbf{Server updates} $c^+$ and $\{c_i^+\}_{i\in g}$ (standard SCAFFOLD \cite{karimireddy2020scaffold})
  \State $A \gets \frac{1}{|g|}\sum_{i\in g} a_i$
\end{algorithmic}
\end{algorithm}

\subsection{Plurality-Vote Inference}
\label{sec:3_vote}
In this section, we describe Step~4, in which the predictions of the trained group models $\{w_g\}_{g\in\mathcal{G}}$ are aggregated via plurality voting (Algorithm~\ref{alg:carvy_vote}). Each model independently produces a predicted label $\hat{y}_g(x)$ and the votes are tallied (lines~3--5). The class with the most votes becomes the final prediction $\hat{y}(x)$ (lines~6--7). In the event of a tie, the class with the smallest index is selected to ensure a deterministic decision rule (line~7).

\begin{algorithm}[t]
\caption{\textsc{PluralityVotePredict} (Step 4): Plurality-vote inference}
\label{alg:carvy_vote}
\begin{algorithmic}[1]
  \Statex \textbf{Input:} trained group models $\{w_g\}_{g \in \mathcal{G}}$; inference input $x$
  \Statex \textbf{Output:} predicted label $\hat{y}(x)$

  \State initialize vote counts $n_b \gets 0$ for all classes $b$
  \For{$g$ \textbf{in} $\mathcal{G}$}
    \State compute logit scores $\{p_{g,b}(x)\}_{b}$ for model $w_g$
    \State $\hat{y}_g(x) \gets \arg\max_{b} p_{g,b}(x)$
    \State $n_{\hat{y}_g(x)} \gets n_{\hat{y}_g(x)} + 1$
  \EndFor
  \State $n_{\max} \gets \max_{b} n_b$
  \State $\hat{y}(x) \gets \min\{\,b \mid n_b = n_{\max}\,\}$
\end{algorithmic}
\end{algorithm}

\section{Evaluation}\label{chap:simulation}
In this section, we evaluate the effectiveness of the proposed method CARVY-FL in a class-disjoint environment, where data distributions across clients are highly skewed and the class sets of different distribution types do not overlap.
The evaluation metrics are certified accuracy (CA) and attack success rate (ASR).
Additionally, we examine the impact of the number of groups on CA in Appendix~\ref{sec_4:group_num}.

\subsection{Evaluation Setup}\label{sec_4:evaluation_setting}
We conducted a simulation-based evaluation of federated learning consisting of a single central server and $N=100$ clients on a single machine.

\subsubsection{Datasets and Data Distribution}
We use MNIST \cite{lecun1998MNIST} and Fashion-MNIST \cite{Xiao2017FMNIST} (FMNIST) as datasets.
MNIST is normalized with mean $\mu=0.1307$ and standard deviation $\sigma=0.3081$, while FMNIST is normalized with mean $\mu=0.2860$ and standard deviation $\sigma=0.3530$.
Of the original 60{,}000 training images, 54{,}000 are used for training and 6{,}000 for validation, and a uniformly distributed test dataset $\mathcal{D}^{\mathrm{test}}$ is used for evaluation.

The number of local training samples per client is set to 50, 100, 200, and 500.
Under the class-disjoint setting, clients are partitioned into 5 distribution types, each containing 20 clients.
Each client's training and validation data are restricted to only 2 classes, with equal numbers of samples assigned to each class, and no data from other classes.
Specifically, for both MNIST and FMNIST, distribution type 1 holds classes 0 and 1, distribution type 2 holds classes 2 and 3, distribution type 3 holds classes 4 and 5, distribution type 4 holds classes 6 and 7, and distribution type 5 holds classes 8 and 9, with equal numbers of samples per class.

\subsubsection{Model}
We use a compact CNN shared across MNIST and FMNIST.
It consists of two $3\times 3$ convolutional layers followed by MaxPool and Dropout, after which the features are flattened and passed through a fully connected layer to produce 10-class outputs. ReLU is used as the activation function.

\subsubsection{Training Conditions and Attacker Behavior}
Training is performed synchronously, where one round consists of the server distributing the model and each client sending back the trained model.
The number of rounds is 200 for MNIST and 300 for FMNIST, and each client performs 1 epoch of training per round.

To address non-IID data, all methods including the baselines employ SCAFFOLD \cite{karimireddy2020scaffold}. The local learning rate is decayed using cosine annealing from an initial value of 0.01 to a minimum value of 0.0001. The global step size is set equal to the number of clients in a group so that the aggregation becomes an equal-weight average of models. The mini-batch SGD batch size is 16, and cross-entropy is used as the loss function.
Early stopping with patience $P=10$ and threshold $\varepsilon=10^{-4}$ is applied based on the mean validation accuracy across clients in each group.

Attackers perform a backdoor attack combining BadNets \cite{gu2017badnets} and model replacement \cite{bagdasaryan2020backdoor}.
The target label is 0, corresponding to digit 0 for MNIST and T-shirt/top for FMNIST.
The attacker takes a fixed subset of 100 images from each of the 9 non-target classes, embeds a $3\times 3$ trigger in the bottom-right corner of each image, and sets the pixel values to 1.0 in the normalized input space.
Local training is then performed using only the poisoned data with labels replaced with 0.

The attacker applies model replacement to the updated model $w_{\text{attack}}$ before sending it to the server. The update rule follows Bagdasaryan et al.~\cite{bagdasaryan2020backdoor}:
\begin{eqnarray}
  w_{\text{send}}
  &=& w_t + \frac{N_{\mathrm{local}}}{V}\left(w_{\text{attack}} - w_t\right)
  \label{eq:model_replacement}
\end{eqnarray}
where $w_t$ is the global model at round $t$, $N_{\mathrm{local}}$ is the number of clients in the group, and $V$ is the number of malicious clients in the same group. Equation~\eqref{eq:model_replacement} is designed so that the attacker's update dominates the equal-weight average aggregation within the group.

\subsubsection{Method Configuration}
For distribution type estimation in CARVY-FL, model updates obtained from 1 epoch of training are compressed to 20 dimensions via PCA, and X-means is applied. The X-means implementation uses xmeans from pyclustering \cite{pycl} 0.10.1.2.
The distance metric is Euclidean distance, the tolerance for X-means is set to the default value of 0.001, and the maximum number of clusters is set to 100, equal to the number of clients.

We compare CARVY-FL with three baseline methods. The first is \textbf{FLCert}~\cite{cao2022flcert}, a voting-based FL method based on hash-based random grouping. The second is \textbf{FLCert+clustering}, an oracle variant that replaces random grouping with ideal clustering-based grouping using ground-truth distribution type information. The third is \textbf{single-global-model}, a standard FL baseline that trains a single global model without any group partitioning. The single-global-model baseline is evaluated for both CA and ASR with 3 random seeds and with the number of malicious clients $m \in \{0, 1, 2, 3, 10, 20\}$.

The number of groups is determined internally by CARVY-FL, whereas for the baseline methods we use $G=20$. We first conducted preliminary experiments over 16 different group counts ranging from 1 to 100. The results showed that $G=20$ provides the strongest attack resilience for FLCert, whereas no single group count consistently outperformed the rest for FLCert+clustering. Accordingly, FLCert+clustering is evaluated using the same group count as FLCert.

Because FLCert assigns clients to groups by hashing, the group sizes are not exactly uniform, although each group contains about five clients on average. By contrast, FLCert+clustering produces groups of exactly five clients. In FLCert, the group ID is computed from the 64-bit client ID $\mathrm{ID}_i$ as $\texttt{hash}(\mathrm{ID}_i)$, with one client pre-assigned to each group to prevent empty groups and the remaining clients assigned by modular arithmetic.

\subsubsection{Evaluation Metrics}
In this evaluation, we use certified accuracy (CA) and attack success rate (ASR) as metrics for attack resilience.

CA represents the lower bound on the fraction of test samples for which the plurality-vote final prediction is correct, assuming that at most $m$ malicious clients capable of inducing arbitrary misclassifications exist among the clients. CA is computed by training without any malicious clients and using the resulting group models. For a uniformly distributed test dataset $\mathcal{D}^{\mathrm{test}}$,
\begin{eqnarray}
  \mathrm{CA}(m)
  &=&
  \frac{
    \bigl|\{(x,y)\in\mathcal{D}^{\mathrm{test}} \mid
      \Delta(x) \ge 2m
    \}\bigr|
  }{
    |\mathcal{D}^{\mathrm{test}}|
  }
  \label{eq:ca_empirical}
\end{eqnarray}
where $\Delta(x)$ denotes the vote margin.

ASR represents the fraction of successful backdoor attacks. Let $\tau(\cdot)$ denote the trigger insertion operation on input $x$, and let $\hat{y}_{\mathrm{vote}}(\tau(x))$ denote the final plurality-vote prediction for the triggered input $\tau(x)$.
To compute ASR, samples whose class is not $y_{\mathrm{target}}$ are extracted from the test data $\mathcal{D}^{\mathrm{test}}$ to form the evaluation set $\mathcal{D}^{\mathrm{trg}}$:
\begin{eqnarray}
  \mathrm{ASR}
  &=&
  \frac{
    \bigl|\{(x,y)\in\mathcal{D}^{\mathrm{trg}} \mid
      \hat{y}_{\mathrm{vote}}(\tau(x)) = y_{\mathrm{target}}
    \}\bigr|
  }{
    |\mathcal{D}^{\mathrm{trg}}|
  }
  \label{eq:asr_empirical}
\end{eqnarray}

For ASR evaluation, we adopt an attacker placement that disperses malicious clients across as many distinct groups as possible, which is disadvantageous for CARVY-FL.
On the other hand, CA provides a lower bound against any attacker placement, so FLCert's advantage of preventing malicious clients from choosing their groups is not captured by the CA metric; instead, CA corresponds to a worst-case evaluation.

\subsection{Attack Resilience Evaluation}\label{sec_4:evaluation}
We compare CA and ASR for CARVY-FL and the baseline methods.
For both MNIST and FMNIST, we evaluate with 50, 100, 200, and 500 training samples per client, computing the mean and standard deviation over 10 trials with different random seeds.

\subsubsection{Certified Accuracy Results}\label{sub_sec_4:CA_evaluation}
\begin{figure*}[t]
  \centering
  \begin{subfigure}[t]{0.49\linewidth}
    \centering
    \includegraphics[width=\linewidth]{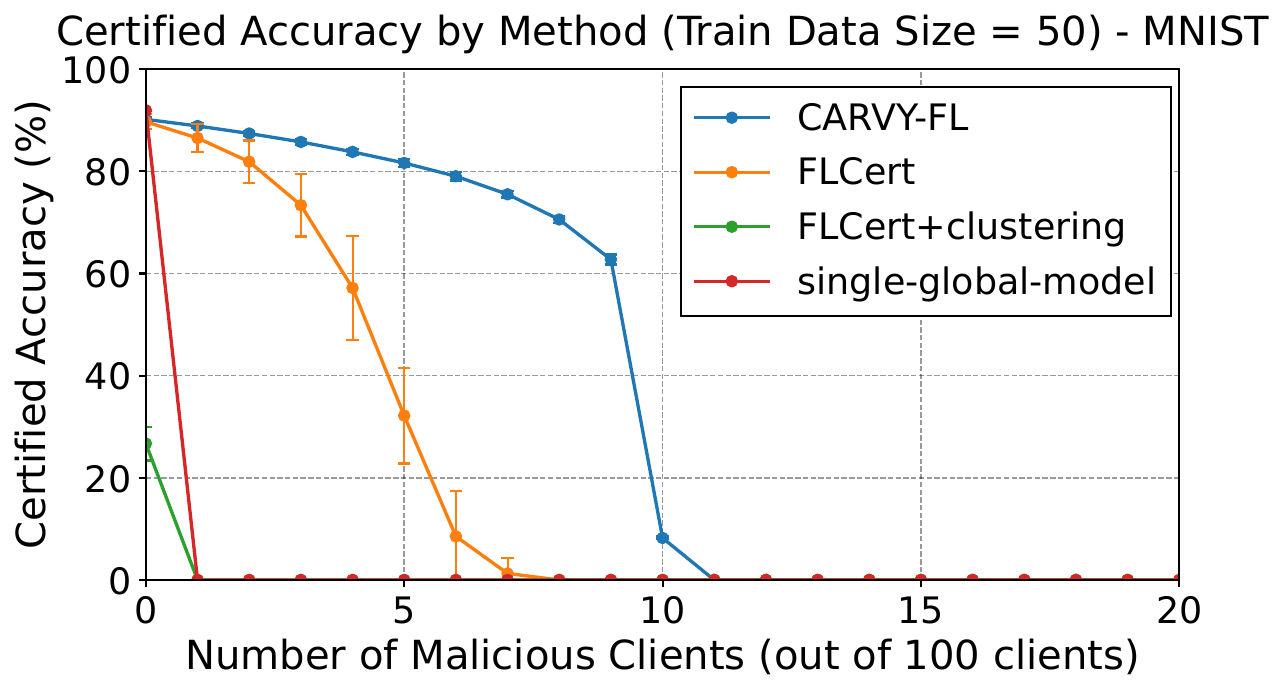}
    \caption{50 training samples}
    \label{fig:MNIST_CA_data50}
  \end{subfigure}
  \hfill
  \begin{subfigure}[t]{0.49\linewidth}
    \centering
    \includegraphics[width=\linewidth]{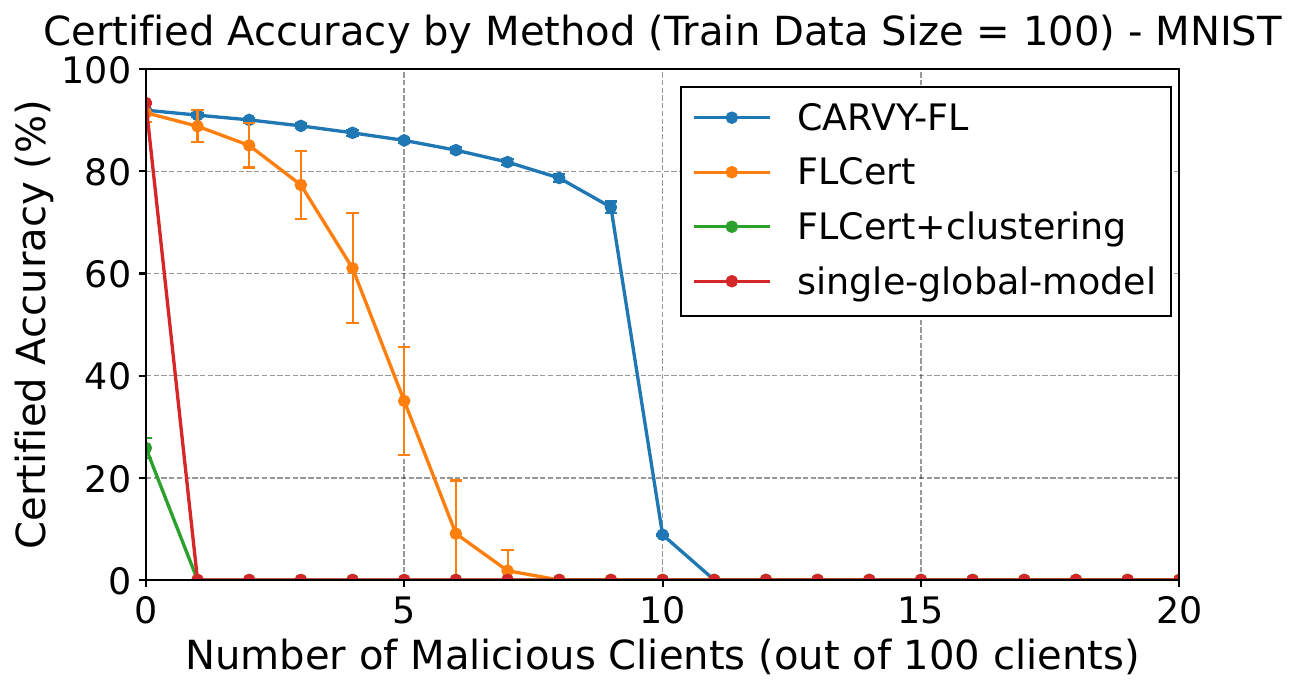}
    \caption{100 training samples}
    \label{fig:MNIST_CA_data100}
  \end{subfigure}

  \begin{subfigure}[t]{0.49\linewidth}
    \centering
    \includegraphics[width=\linewidth]{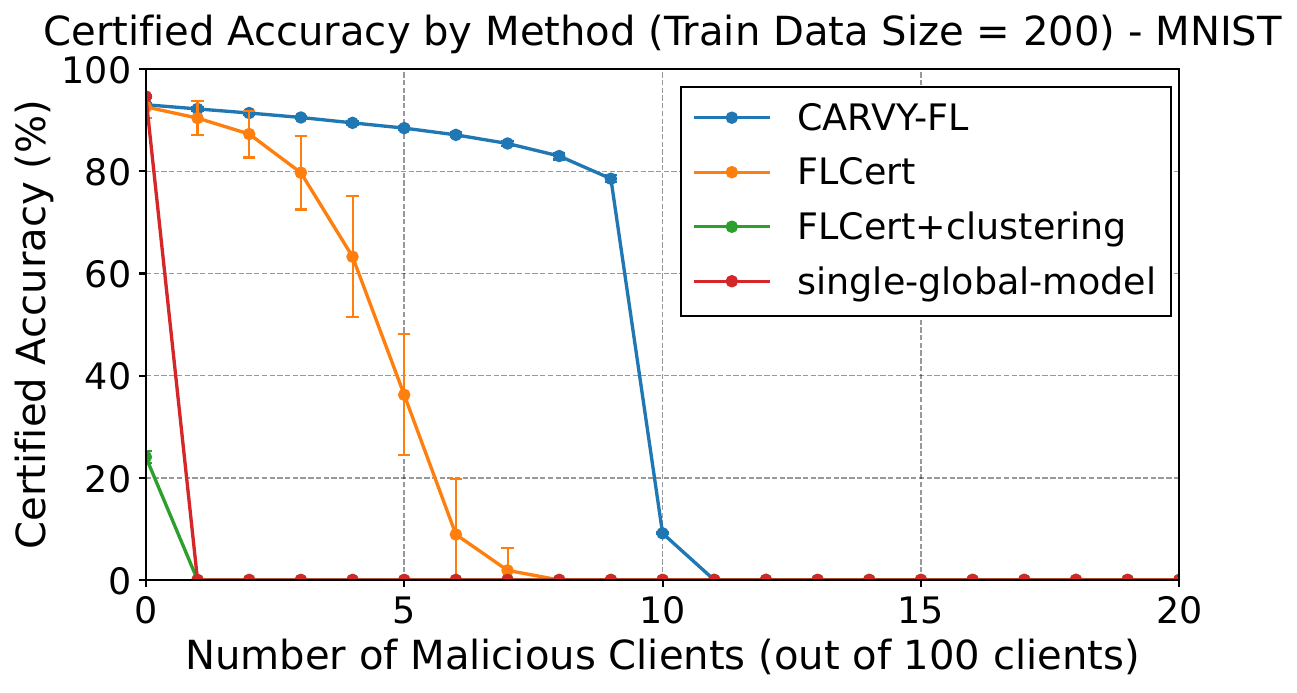}
    \caption{200 training samples}
    \label{fig:MNIST_CA_data200}
  \end{subfigure}
  \hfill
  \begin{subfigure}[t]{0.49\linewidth}
    \centering
    \includegraphics[width=\linewidth]{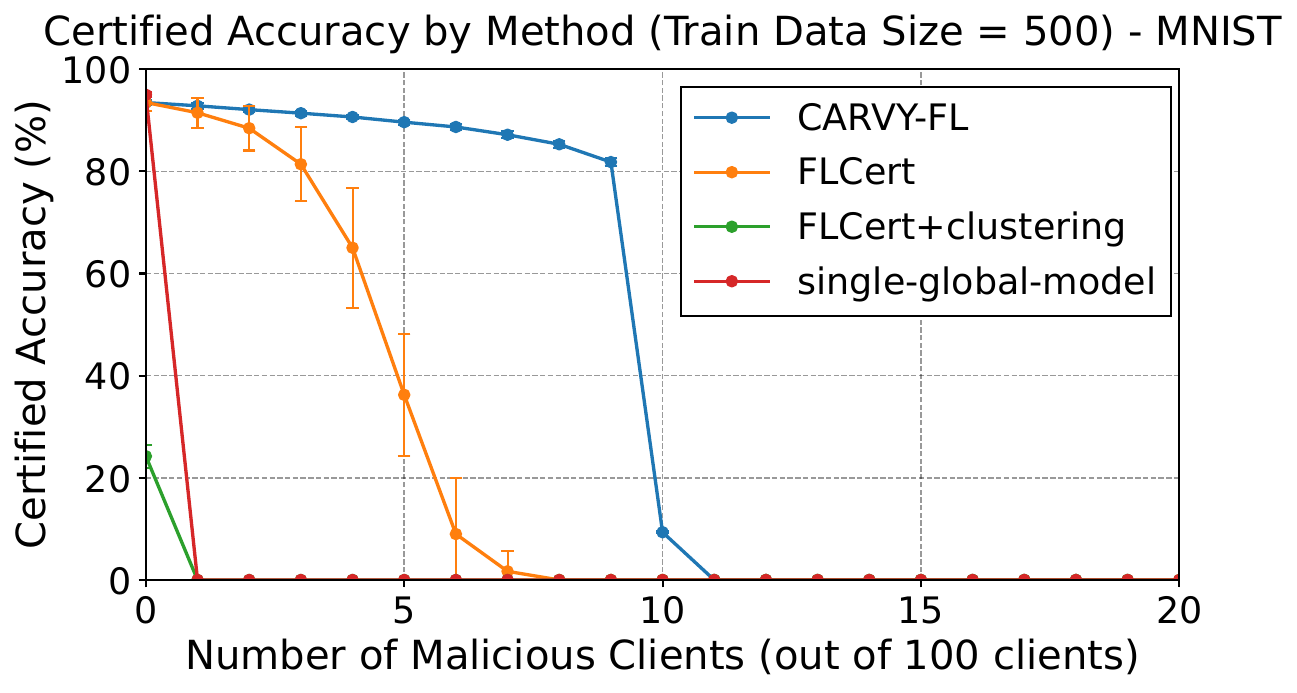}
    \caption{500 training samples}
    \label{fig:MNIST_CA_data500}
  \end{subfigure}
  \caption{Certified accuracy for different training data sizes on MNIST.}
  \Description{Four line plots showing certified accuracy versus number of malicious clients on MNIST for 50, 100, 200, and 500 training samples per client, comparing CARVY-FL, FLCert, and FLCert+clustering.}
  \label{fig:MNIST_CA_datasize_2x2}
\end{figure*}

\begin{figure*}[t]
  \centering
  \begin{subfigure}[t]{0.49\linewidth}
    \centering
    \includegraphics[width=\linewidth]{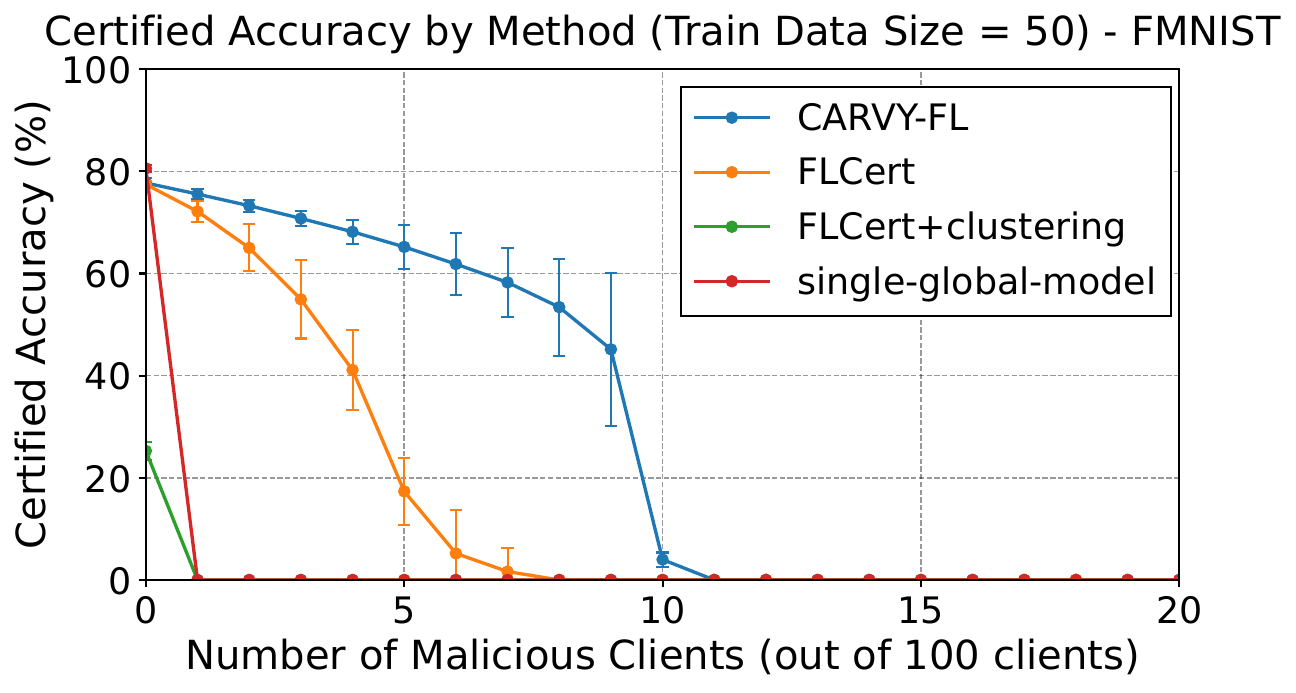}
    \caption{50 training samples}
    \label{fig:FMNIST_CA_data50}
  \end{subfigure}
  \hfill
  \begin{subfigure}[t]{0.49\linewidth}
    \centering
    \includegraphics[width=\linewidth]{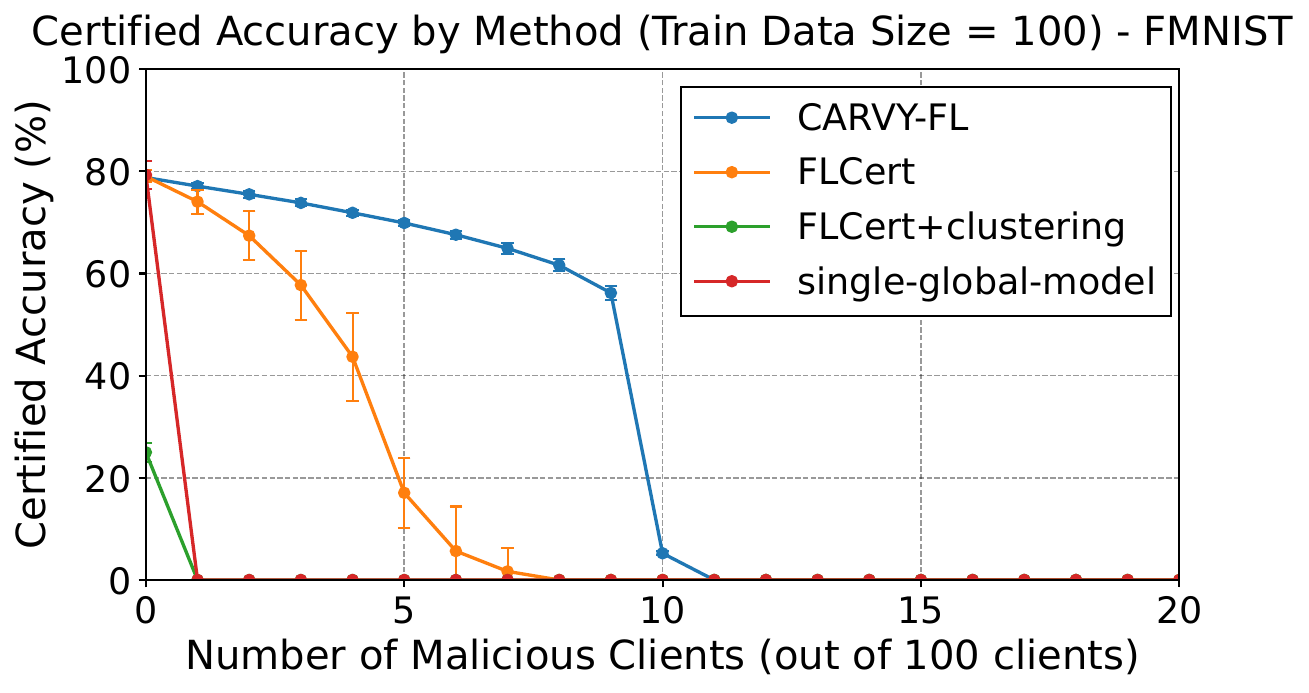}
    \caption{100 training samples}
    \label{fig:FMNIST_CA_data100}
  \end{subfigure}

  \begin{subfigure}[t]{0.49\linewidth}
    \centering
    \includegraphics[width=\linewidth]{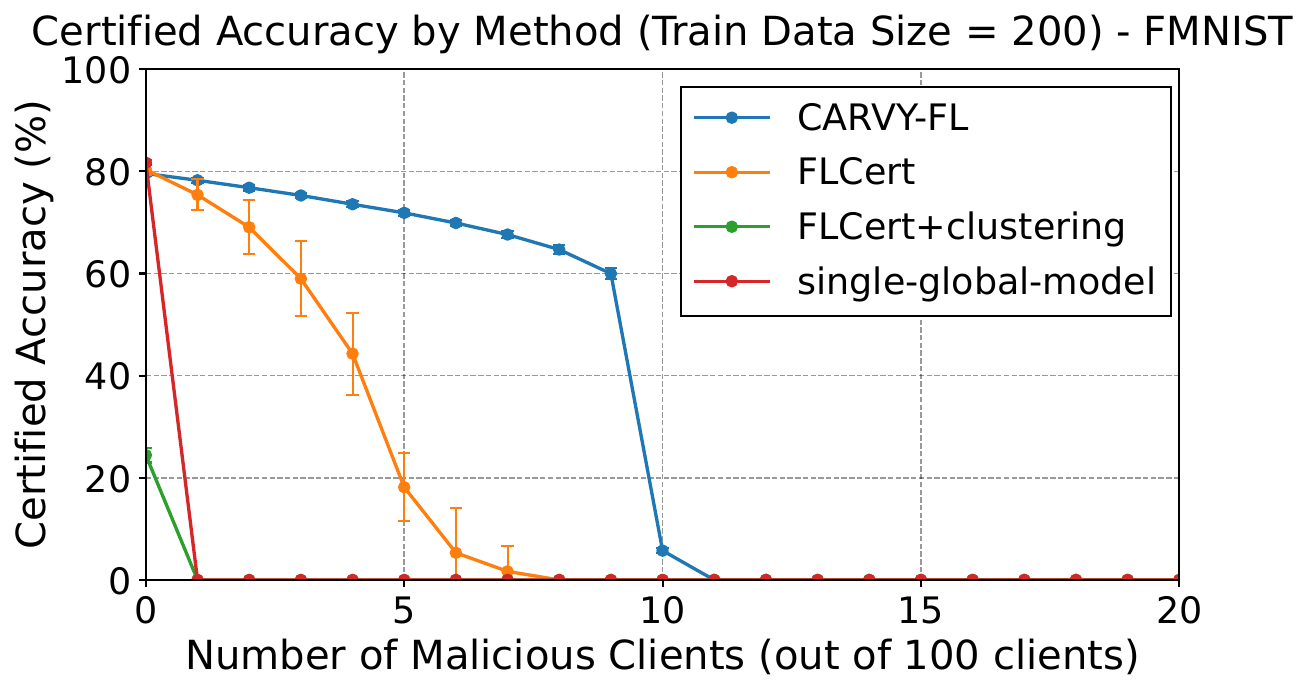}
    \caption{200 training samples}
    \label{fig:FMNIST_CA_data200}
  \end{subfigure}
  \hfill
  \begin{subfigure}[t]{0.49\linewidth}
    \centering
    \includegraphics[width=\linewidth]{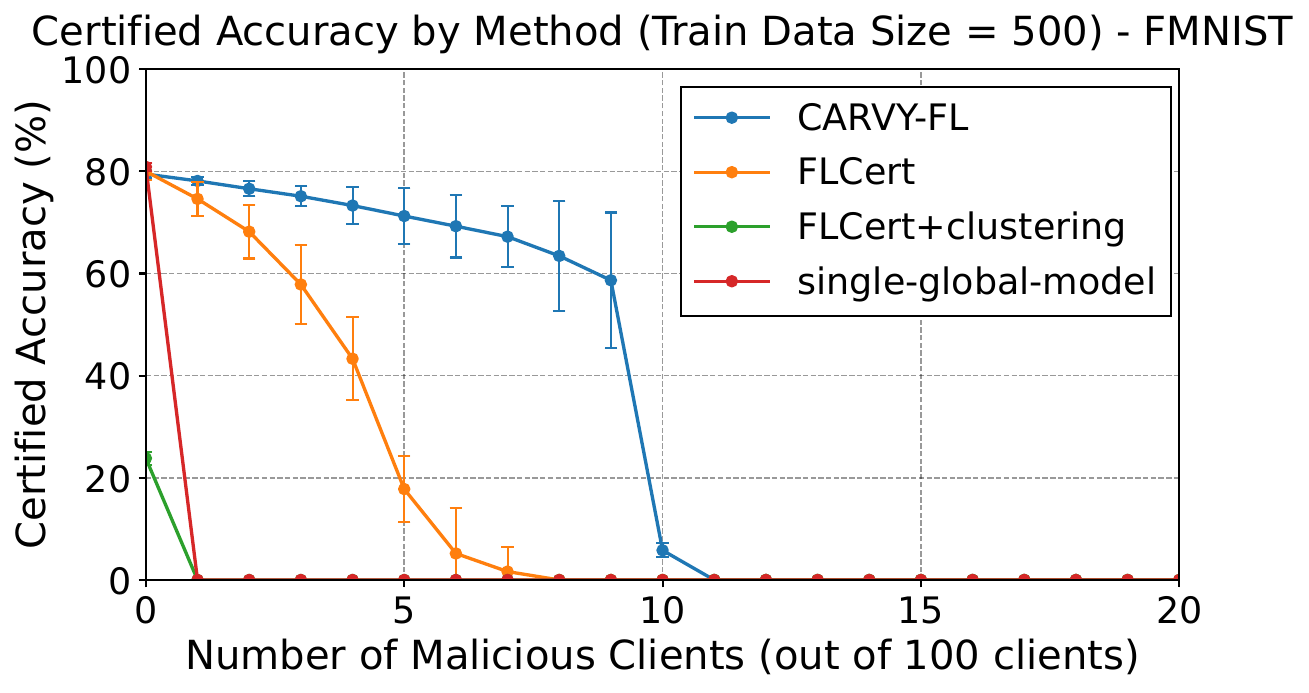}
    \caption{500 training samples}
    \label{fig:FMNIST_CA_data500}
  \end{subfigure}
  \caption{Certified accuracy for different training data sizes on FMNIST.}
  \Description{Four line plots showing certified accuracy versus number of malicious clients on FMNIST for 50, 100, 200, and 500 training samples per client, comparing CARVY-FL, FLCert, and FLCert+clustering.}
  \label{fig:FMNIST_CA_datasize_2x2}
\end{figure*}

Figure~\ref{fig:MNIST_CA_datasize_2x2} and Figure~\ref{fig:FMNIST_CA_datasize_2x2} show CA as a function of the number of malicious clients $m$.

The single-global-model baseline, which represents standard FL without group partitioning, shows no attack resilience: its CA drops to nearly 0\% with even a single malicious client ($m=1$) on both MNIST and FMNIST across all training data sizes. This confirms that conventional FL is highly vulnerable to poisoning attacks and motivates the need for group-based certified defenses.

For MNIST with 50 training samples, CARVY-FL achieves $\mathrm{CA}(0)$ of $90.2$\%, which is comparable to FLCert's $89.7$\% and approximately 3.4 times higher than FLCert+clustering's $26.7$\%.
The maximum number of malicious clients for which CA exceeds 80\% is $m=5$ for CARVY-FL and $m=2$ for FLCert.
The maximum number of malicious clients for which CA exceeds 60\%, 40\%, and 20\% is $m=9$ for CARVY-FL in all cases, and $m=3,4,5$ for FLCert, respectively.
The standard deviation of CARVY-FL is within approximately 1\% across all measured malicious client counts, whereas FLCert exhibits a standard deviation of around 10\% at $m=4$; CARVY-FL's standard deviation is roughly one-tenth of that.
This trend holds for training data sizes of 100, 200, and 500 as well, with the number of groups in CARVY-FL being 20 in all cases.

For FMNIST, CARVY-FL maintains comparable certified accuracy while tolerating up to approximately twice as many malicious clients as FLCert.
However, with 50 and 500 training samples, the standard deviation of CARVY-FL increases and exceeds that of FLCert. Specifically, the maximum standard deviation of CARVY-FL is 15.0 for 50 training samples and 13.3 for 500 training samples, while for FLCert, the corresponding values are 8.54 and 8.95, respectively. A discussion of this result is provided in Section~\ref{sec:consideration}.

\subsubsection{Distribution Type Estimation Results}
\begin{figure}[t]
  \centering
  \includegraphics[width=1\linewidth]{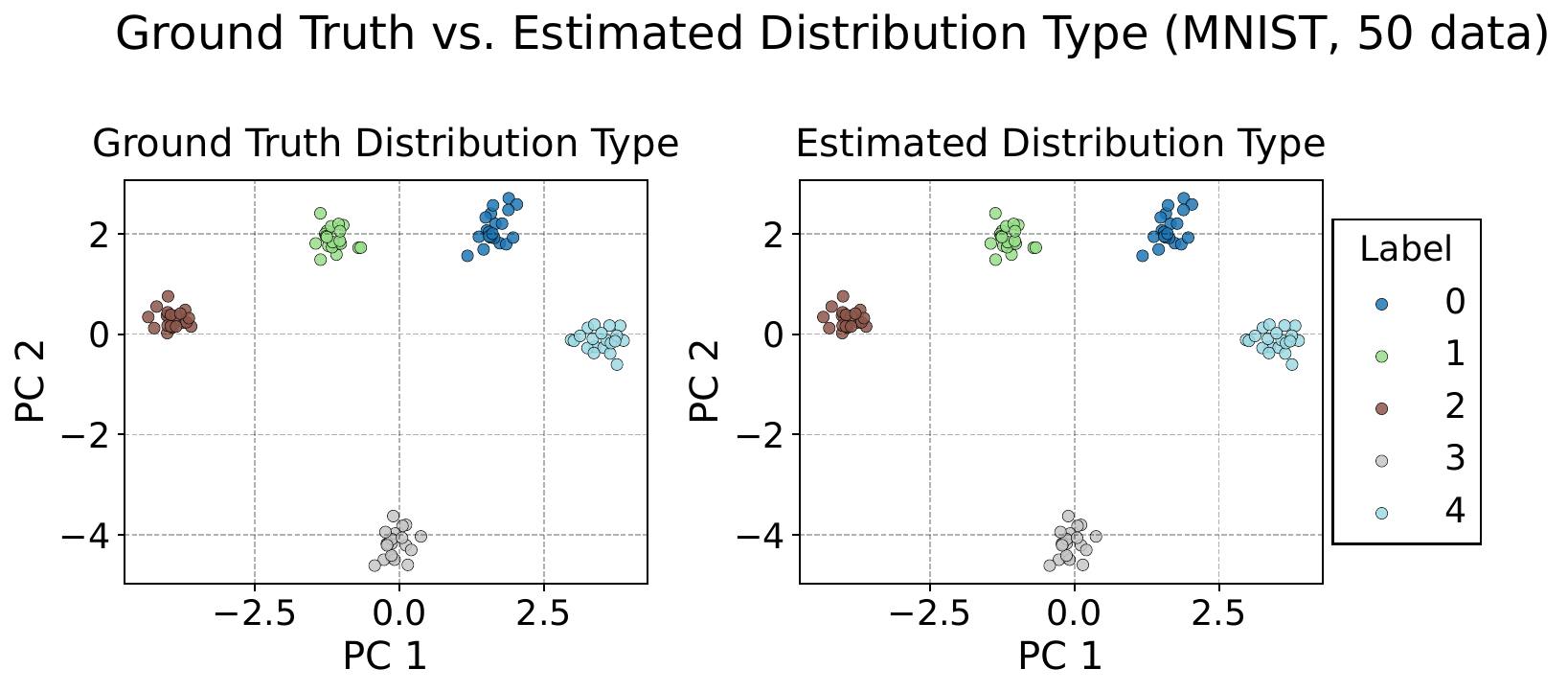}
  \caption{Distribution type estimation results on MNIST. 50 training samples.}
  \Description{Scatter plot of PCA-projected model updates on MNIST with 50 training samples, showing five distinct clusters corresponding to distribution types.}
  \label{fig:MNIST_grouping_data50}
\end{figure}

\begin{figure}[t]
  \centering
  \includegraphics[width=1\linewidth]{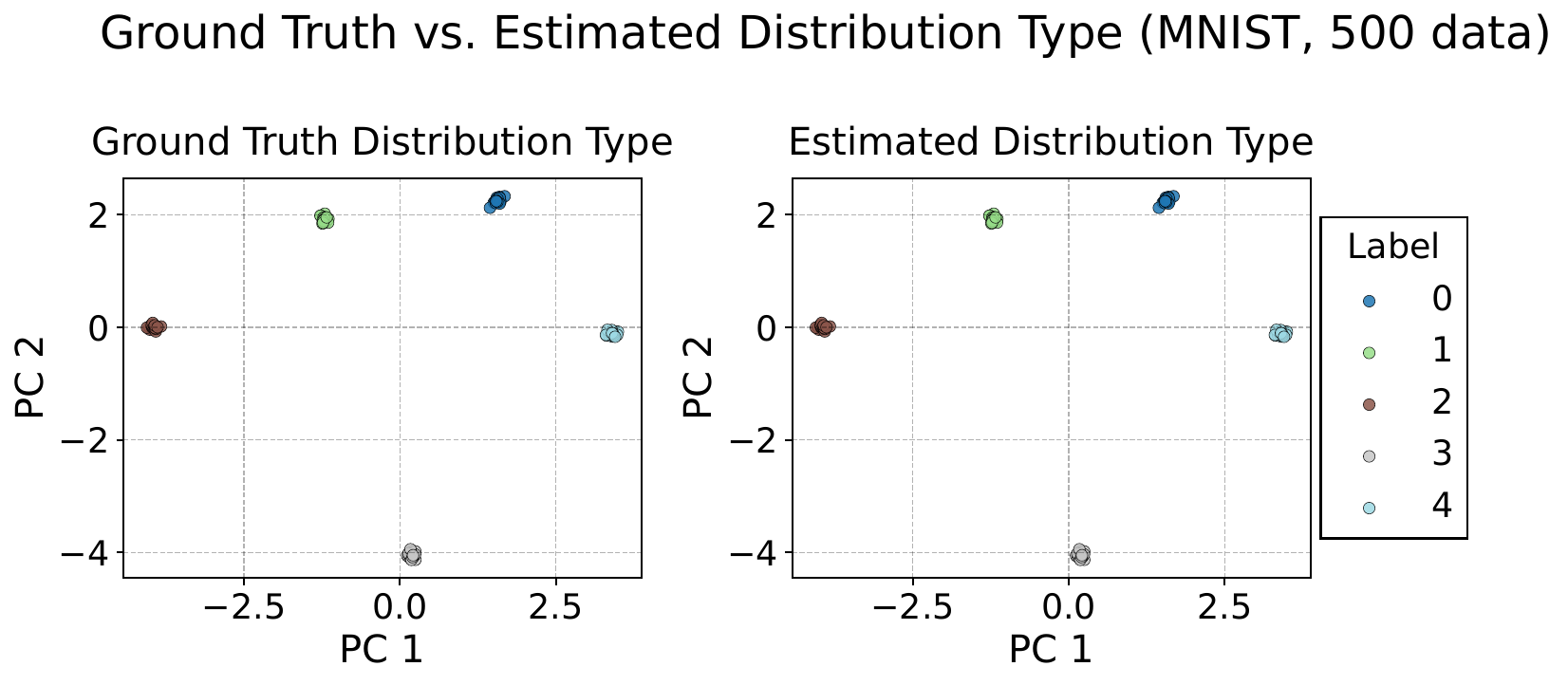}
  \caption{Distribution type estimation results on MNIST. 500 training samples.}
  \Description{Scatter plot of PCA-projected model updates on MNIST with 500 training samples, showing five distinct clusters corresponding to distribution types.}
  \label{fig:MNIST_grouping_data500}
\end{figure}

\begin{figure}[t]
  \centering
  \includegraphics[width=1\linewidth]{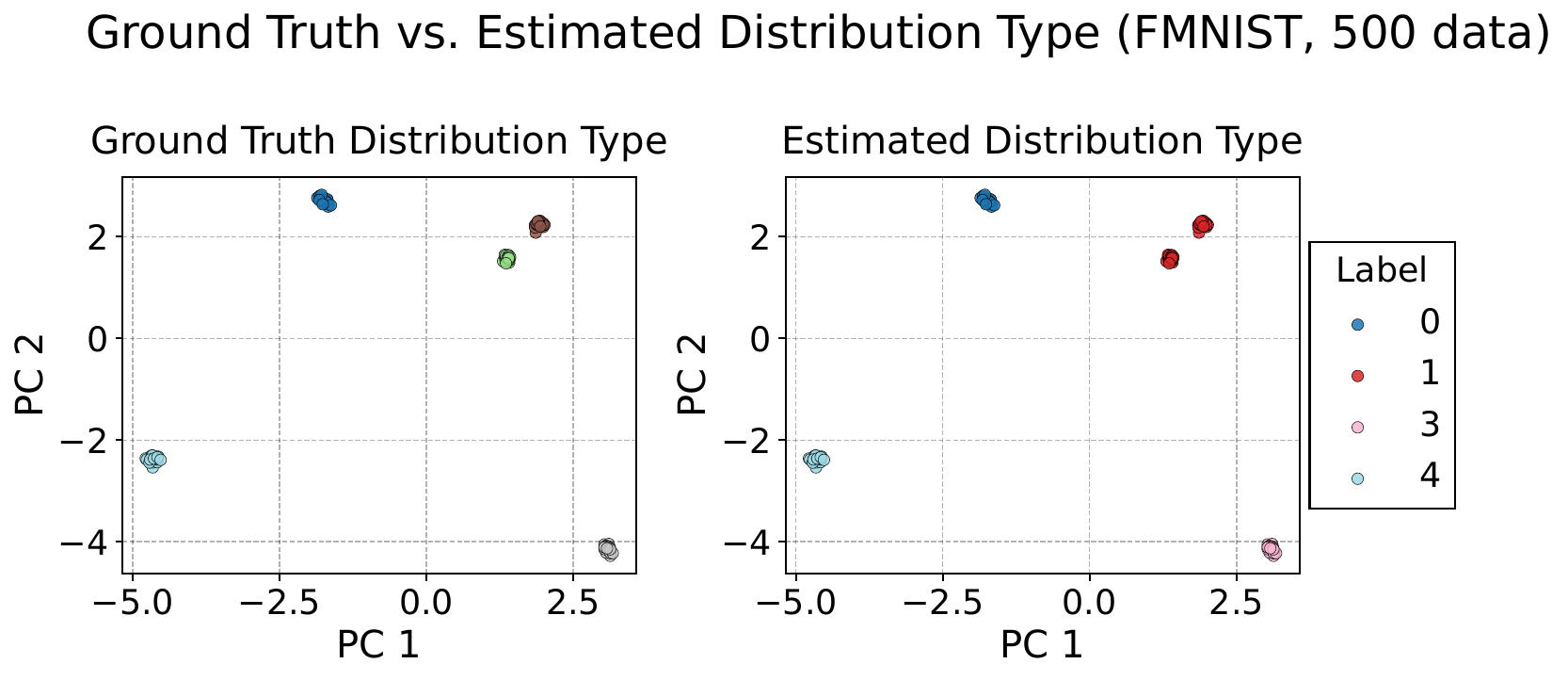}
  \caption{Distribution type estimation results on FMNIST. 500 training samples. Multiple distribution types are conflated into the same cluster.}
  \Description{Scatter plot of PCA-projected model updates on FMNIST with 500 training samples, showing overlapping clusters where multiple distribution types are conflated.}
  \label{fig:FMNIST_grouping_data500}
\end{figure}

Visualizations of the distribution type estimation are shown in Figure~\ref{fig:MNIST_grouping_data50}, Figure~\ref{fig:MNIST_grouping_data500}, and Figure~\ref{fig:FMNIST_grouping_data500}.
Each figure is a scatter plot of model updates obtained from 1 epoch of training projected onto 2 dimensions via PCA.

For MNIST, distribution type estimation succeeded in all trials.
For FMNIST, estimation succeeded in all trials with 100 and 200 training samples, yielding 20 groups based on the estimation results.
With 50 and 500 training samples, X-means under-segmentation occurred in 1 out of 10 trials each. In these failure cases, the number of groups was 25.

\subsubsection{ASR Results Under Actual Attacks}
\begin{figure}[t]
  \centering
  \begin{subfigure}[t]{\linewidth}
    \centering
    \includegraphics[width=\linewidth]{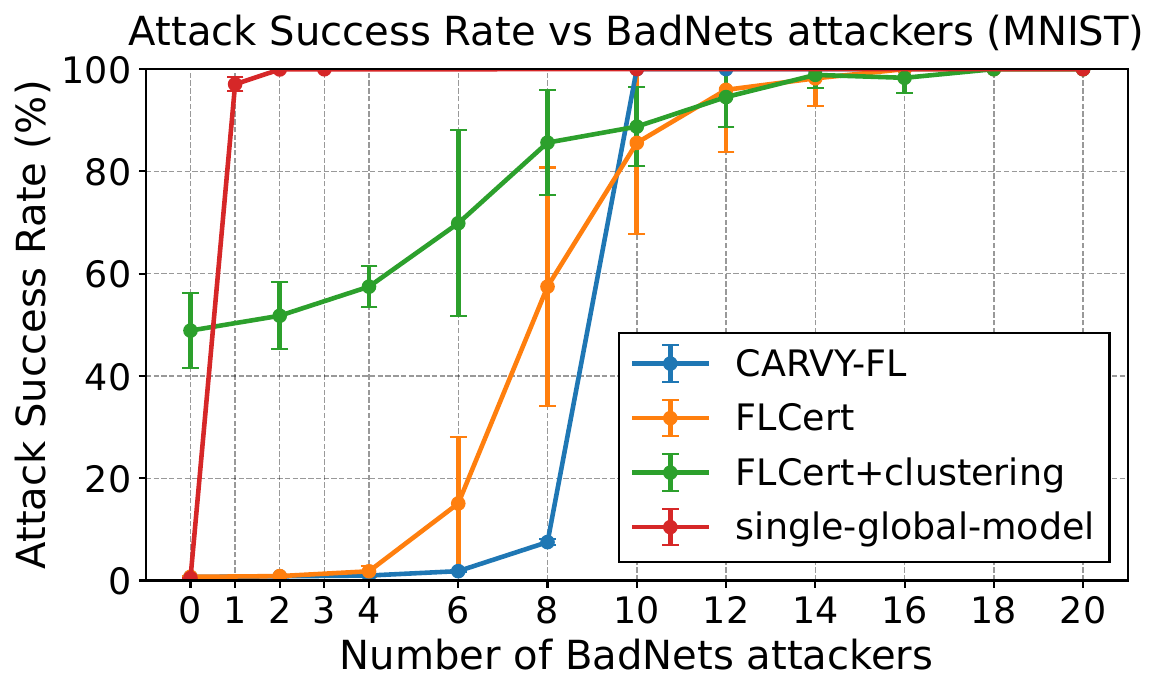}
    \caption{MNIST}
    \label{fig:MNIST_ASR}
  \end{subfigure}

  \begin{subfigure}[t]{\linewidth}
    \centering
    \includegraphics[width=\linewidth]{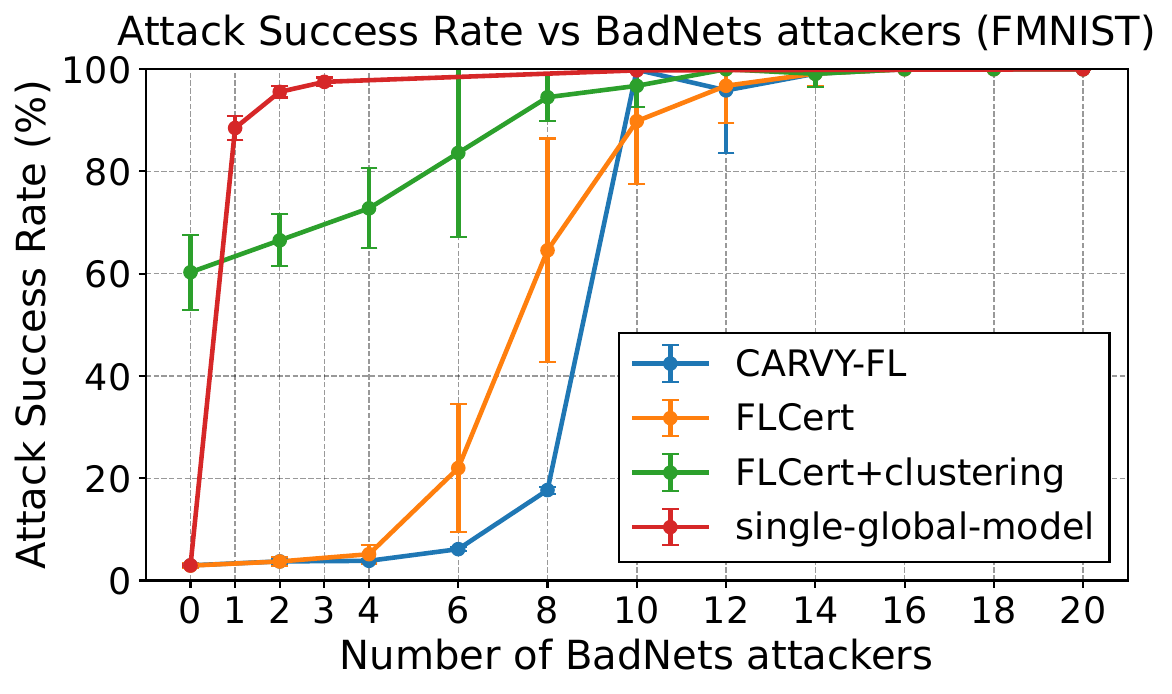}
    \caption{FMNIST}
    \label{fig:FMNIST_ASR}
  \end{subfigure}
  \caption{Attack success rate under actual attacks using BadNets and model replacement.}
  \Description{Two line plots showing attack success rate versus number of malicious clients on MNIST and FMNIST with 500 training samples, comparing CARVY-FL, FLCert, and FLCert+clustering.}
  \label{fig:ASR}
\end{figure}

Figure~\ref{fig:ASR} shows the ASR evaluation results with 500 training samples.
The single-global-model baseline reaches an ASR of nearly 100\% with only 1--3 malicious clients on both MNIST and FMNIST, again demonstrating the lack of attack resilience in standard FL.

Compared to FLCert, CARVY-FL reduces the mean ASR at $m=6$ by approximately 15\% on MNIST and approximately 10\% on FMNIST.
At $m=8$, the reduction is approximately 50\% on MNIST and approximately 40\% on FMNIST.
On the other hand, when the number of malicious clients reaches or exceeds half the number of groups, i.e., $m=10,11$, the ASR of CARVY-FL increases in some cases.

To summarize the ASR trends, Table~\ref{tab:asr_auc_m0} shows the Area Under the Curve (AUC) for $100-\mathrm{ASR}$. The AUC of $100-\mathrm{ASR}$ in this evaluation represents the area above the curve in a plot with ASR on the vertical axis and the number of malicious clients on the horizontal axis.
The AUC of CARVY-FL is 877.00 on MNIST and 844.58 on FMNIST, surpassing FLCert's 789.36 and 734.84.
The improvement rate is 11.1\% on MNIST and 14.9\% on FMNIST.
Moreover, compared to FLCert+clustering, the increase is 143\% on MNIST and 296\% on FMNIST.

\begin{table}[b]
  \centering
  \caption{AUC of $100-\mathrm{ASR}$ $\uparrow$.}
  \label{tab:asr_auc_m0}
  \footnotesize
  \renewcommand{\arraystretch}{1.10}
  \begin{tabular}{l r r}
    \toprule
    Method & MNIST & FMNIST \\
    \midrule
    CARVY-FL & \textbf{877.00} & \textbf{844.58} \\
    FLCert & 789.36 & 734.84 \\
    FLCert+clustering & 360.69 & 213.24 \\
    \bottomrule
  \end{tabular}
\end{table}

\subsubsection{Discussion}\label{sec:consideration}
The CA results demonstrate that CARVY-FL can maintain a high accuracy lower bound across a wide range of malicious client counts in a class-disjoint environment.
For MNIST with 50 training samples, the maximum number of malicious clients for which CA exceeds 80\% is $m=5$ for CARVY-FL and $m=2$ for FLCert, indicating that CARVY-FL provides comparable or superior certified accuracy while tolerating approximately twice as many malicious clients.

\begin{figure}[t]
  \centering
  \includegraphics[width=1\linewidth]{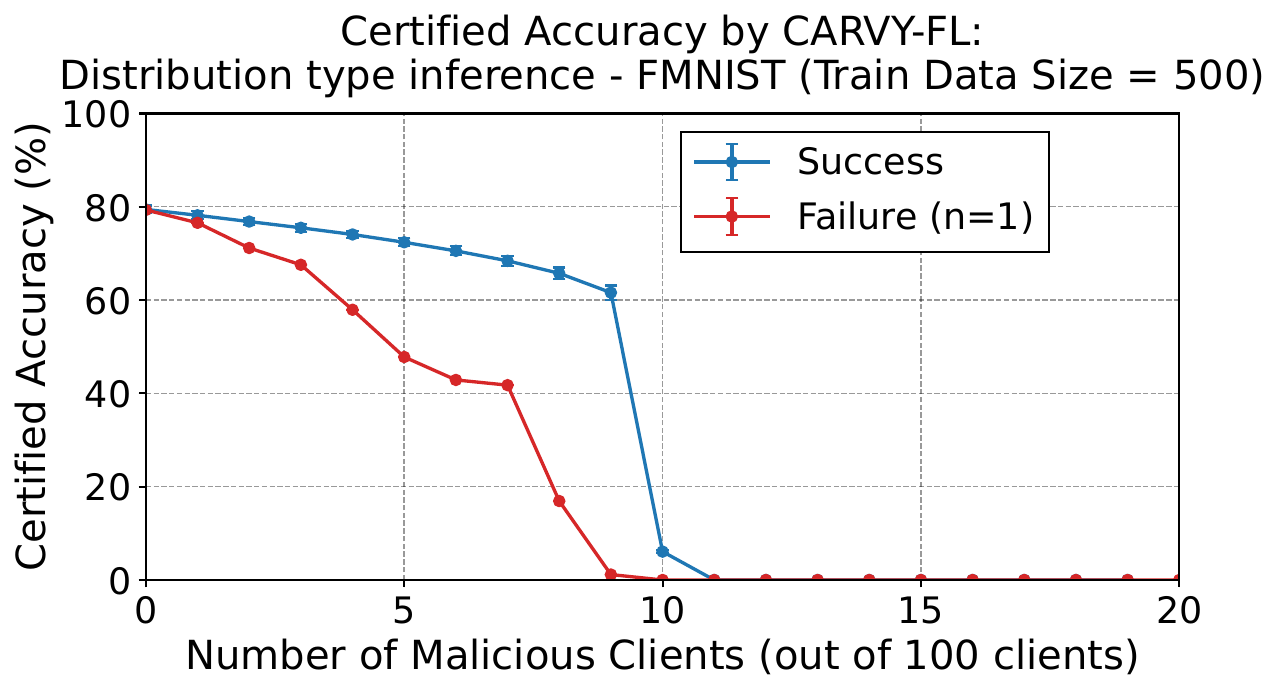}
  \caption{Certified accuracy on FMNIST with 500 training samples when CARVY-FL's distribution type estimation succeeds versus when it fails.}
  \Description{Line plot comparing certified accuracy on FMNIST with 500 training samples for successful versus failed distribution type estimation trials in CARVY-FL.}
  \label{fig:miss_grouping_data500}
\end{figure}

When distribution type estimation succeeds, CARVY-FL forms 20 groups and uses Anticlustering to evenly distribute the 5 distribution types within each group.
This configuration promotes generalization of each group model, increases the vote margin for the correct class in inter-group plurality voting, and thereby raises CA overall.
For FMNIST with 50 and 500 training samples, distribution type estimation failed in 1 out of 10 trials, resulting in 25 groups. As shown in Figure~\ref{fig:miss_grouping_data500}, these failed trials lower the mean CA and increase the variance. Nevertheless, even in the failed trials, CARVY-FL still achieves higher CA than FLCert across all malicious client counts, indicating that the proposed method remains effective even when distribution type estimation is imperfect.

Regarding ASR, CARVY-FL is advantageous at moderate malicious client counts: with 500 training samples, it reduces ASR by approximately 15\% on MNIST and approximately 10\% on FMNIST at $m=6$, and by approximately 50\% on MNIST and approximately 40\% on FMNIST at $m=8$.
On the other hand, CARVY-FL is disadvantaged in some cases at $m=10,11$. In this evaluation, malicious clients are dispersed across different groups for CARVY-FL, meaning that 10 malicious clients can dominate 10 groups at $m=10$.
For FLCert, malicious clients are assigned by a hash function, and at $m=10$, the number of groups containing malicious clients is $8.7\pm 0.64$ on average, remaining below half.
This difference manifests around $m=10,11$.
Nevertheless, overall, the AUC of $100-\mathrm{ASR}$ for CARVY-FL is 877.00 and 844.58, exceeding FLCert's 789.36 and 734.84, confirming that the proposed method achieves superior attack resilience.

\section{Conclusion}
\label{sec:conclusion}

We proposed CARVY-FL, a voting-based federated learning method that replaces random client grouping with Anticlustering-based grouping to improve robustness against poisoning attacks in class-disjoint non-IID settings. CARVY-FL estimates client distribution types from single-epoch model updates, constructs groups with maximal distributional diversity, trains per-group models via SCAFFOLD, and aggregates predictions by plurality voting.

Experiments on MNIST and FMNIST demonstrated that a standard single-global-model FL baseline offers no attack resilience, with certified accuracy dropping to nearly 0\% when even a single malicious client is present and attack success rate reaching nearly 100\% with only 1--3 malicious clients. In contrast, CARVY-FL tolerates approximately twice as many malicious clients as FLCert while maintaining comparable certified accuracy. For ASR under BadNets with model replacement, CARVY-FL improved the AUC of $100{-}\mathrm{ASR}$ by 11.1\% on MNIST and 14.9\% on FMNIST over FLCert.

Future work includes extending the evaluation to more complex datasets and models, and broadening the scope beyond class-disjoint settings to other non-IID distributions such as class imbalance and feature shifts. Another direction is designing grouping strategies that maximize attack resilience under a budget constraint on the number of group models.

\bibliographystyle{ACM-Reference-Format}
\bibliography{references}

\appendix
\section{CA Under Varying Numbers of Groups}\label{sec_4:group_num}
Certified accuracy is significantly affected not only by the diversity of clients within each group but also by the number of groups.
Since the plurality voting mechanism adopted in this work casts one vote per group, the number of groups equals the total number of votes.
Increasing the number of groups raises the count of those composed entirely of benign clients for a given number of malicious clients, thereby increasing the number of unaffected votes.
On the other hand, a larger number of groups reduces the average number of clients per group, which tends to degrade the federated learning accuracy within each group.
Given this trade-off, CARVY-FL automatically determines the number of groups.
We therefore verify that the group count determination logic of CARVY-FL is effective, and that the group count of 20 used in the previous subsection is appropriate for FLCert and FLCert+clustering.

With 100 clients, we varied the number of groups from $1$ to $100$ for CARVY-FL, FLCert, and FLCert+clustering.
For CARVY-FL and FLCert+clustering, clients were allocated to groups as evenly as possible, with the difference between any two groups being at most $\pm 1$.
In CARVY-FL, when the number of clients per group is fewer than the number of distribution types (5), i.e., when the number of groups exceeds 20, clients are assigned so that no distribution type overlaps within a group.
Conversely, when the number of clients per group exceeds 5, i.e., when the number of groups is fewer than 20, each group contains clients from all distribution types in roughly equal proportions with a difference of at most $\pm 1$.
For FLCert, clients were assigned to groups based on a hash function with the specified number of groups $G$.

The evaluation settings are shown in Table~\ref{tab:mnist_setting_groupnum}.
The datasets used are MNIST and FMNIST, with 500 training samples per client.
The number of trials is 3 for CARVY-FL and FLCert+clustering (which exhibit low variance) and 6 for FLCert (which exhibits high variance).
Only the mean values are plotted; standard deviations are omitted.
All other settings follow Section~\ref{sec_4:evaluation}.

\begin{table}[b]
  \centering
  \caption{Evaluation settings (group count comparison).}
  \label{tab:mnist_setting_groupnum}
  \begin{tabular}{ll}
    \toprule
    Item & Setting \\
    \midrule
    Dataset & MNIST, FMNIST \\
    Training samples per client & 500 \\
    Number of groups & 1, 2, 4, 5, 8, \\ & 10, 11, 12, 14, 16, \\ & 18, 20, 22, 25, 33,\\ & 50, 80, 100 \\
    Number of trials & 6 for FLCert; 3 for others \\
    Other settings & Same as Section~\ref{sec_4:evaluation} \\
    \bottomrule
  \end{tabular}
\end{table}

\begin{table}[b]
  \centering
  \caption{AUC of certified accuracy for each number of groups on MNIST $\uparrow$.}
  \label{tab:groupnum_auc_mnist}
  \small
  \setlength{\tabcolsep}{4pt}
  \begin{tabular}{r rrr}
    \toprule
    G & CARVY-FL & FLCert & FLCert+clust. \\
    \midrule
     1  &  47.25 &  47.32 & \textbf{47.52} \\
     2  &  57.01 &  56.82 &  29.34         \\
     4  & 148.82 & 143.82 &  32.24         \\
     5  & 230.07 & 221.17 &   9.99         \\
     8  & 328.07 & 307.73 &  28.09         \\
    10  & 418.59 & 334.65 &  11.18         \\
    11  & 491.47 & 344.18 &  35.39         \\
    12  & 499.59 & 389.86 &  29.73         \\
    14  & 586.89 & 403.48 &  31.43         \\
    16  & 673.99 & 389.66 &  22.97         \\
    18  & 765.71 & 374.42 &  27.91         \\
    20  & \textbf{852.67} & \textbf{413.08} &  12.28 \\
    22  & 818.92 & 405.85 &  11.64         \\
    25  & 706.49 & 371.99 &  12.65         \\
    33  & 511.08 & 356.49 &  29.21         \\
    50  & 219.95 & 245.11 &  15.18         \\
    80  &  27.63 &  95.15 &  24.46         \\
   100  &  21.37 &  19.30 &  19.00         \\
    \bottomrule
  \end{tabular}

  \footnotesize
  Note: For $G{=}1$ and $G{=}100$, CARVY-FL and FLCert+clustering ideally yield identical results; however, differences in grouping logic cause variations in training order, leading to slight numerical discrepancies.
\end{table}

\begin{table}[b]
  \centering
  \caption{AUC of certified accuracy for each number of groups on Fashion-MNIST $\uparrow$.}
  \label{tab:groupnum_auc_fmnist}
  \small
  \setlength{\tabcolsep}{4pt}
  \begin{tabular}{r rrr}
    \toprule
    G & CARVY-FL & FLCert & FLCert+clust. \\
    \midrule
     1  &  40.49 &  40.37 & \textbf{40.51} \\
     2  &  47.87 &  47.28 &  26.66         \\
     4  & 124.05 & 109.29 &  25.79         \\
     5  & 189.33 & 166.99 &   9.87         \\
     8  & 260.49 & 229.08 &  19.97         \\
    10  & 341.30 & 254.91 &  10.89         \\
    11  & 384.03 & 274.31 &  35.39         \\
    12  & 387.74 & 282.85 &  26.18         \\
    14  & 443.83 & 300.33 &  25.64         \\
    16  & 518.95 & 280.50 &  24.25         \\
    18  & 599.47 & 255.91 &  25.45         \\
    20  & \textbf{686.02} & \textbf{300.87} &  12.05 \\
    22  & 646.35 & 289.23 &  11.46         \\
    25  & 550.47 & 276.23 &  12.27         \\
    33  & 372.82 & 277.06 &  18.80         \\
    50  & 116.37 & 207.09 &  14.59         \\
    80  &  20.72 &  73.47 &  16.24         \\
   100  &  17.43 &  17.86 &  17.59         \\
    \bottomrule
  \end{tabular}

  \footnotesize
  Note: For $G{=}1$ and $G{=}100$, CARVY-FL and FLCert+clustering ideally yield identical results; however, differences in grouping logic cause variations in training order, leading to slight numerical discrepancies.
\end{table}

\begin{figure*}[t]
  \centering

  \begin{subfigure}[t]{0.49\linewidth}
    \centering
    \includegraphics[width=\linewidth]{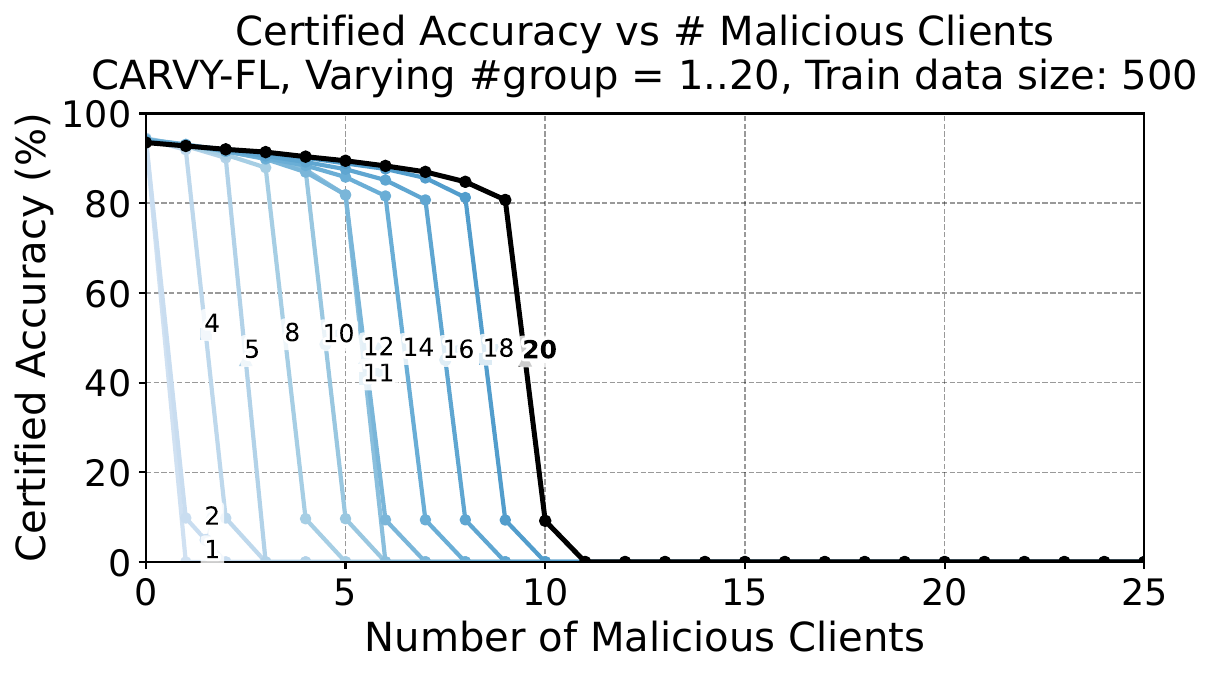}
    \caption{MNIST: $1 \le G \le 20$}
    \label{fig:MNIST_group_num_CARVY-FL_20under}
  \end{subfigure}\hfill
  \begin{subfigure}[t]{0.49\linewidth}
    \centering
    \includegraphics[width=\linewidth]{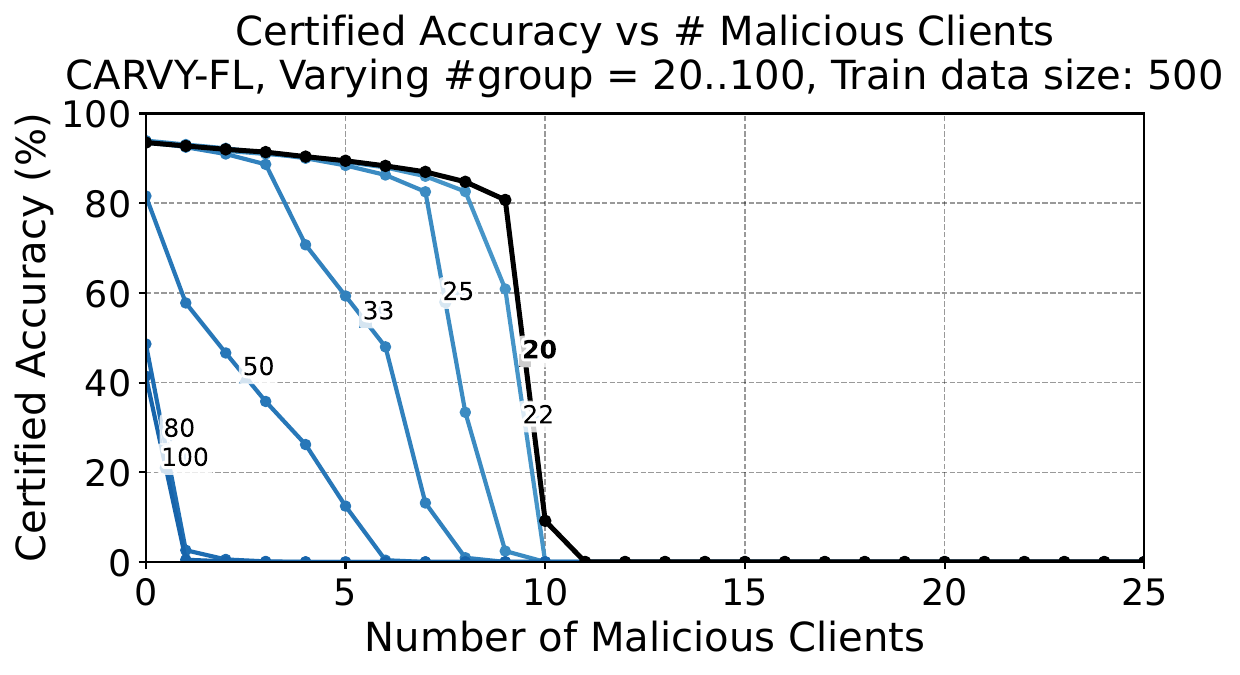}
    \caption{MNIST: $20 \le G \le 100$}
    \label{fig:MNIST_group_num_CARVY-FL_20over}
  \end{subfigure}

  \begin{subfigure}[t]{0.49\linewidth}
    \centering
    \includegraphics[width=\linewidth]{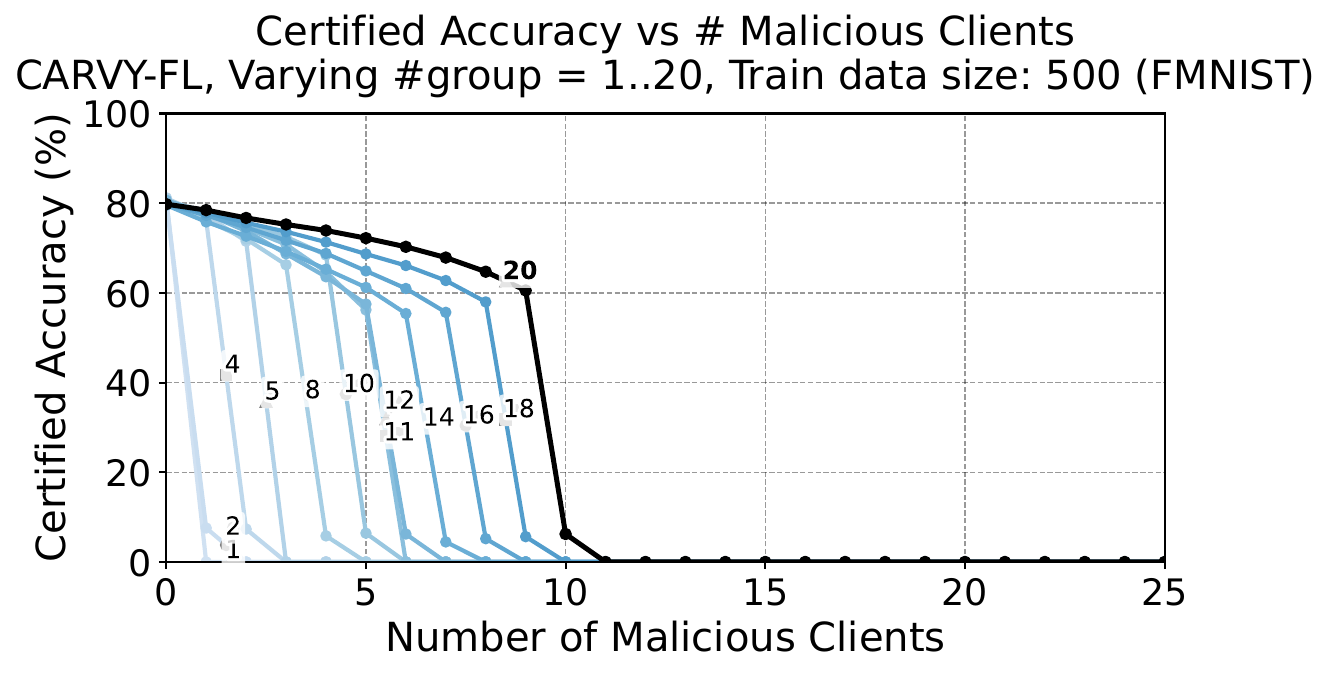}
    \caption{FMNIST: $1 \le G \le 20$}
    \label{fig:FMNIST_group_num_CARVY-FL_20under}
  \end{subfigure}\hfill
  \begin{subfigure}[t]{0.49\linewidth}
    \centering
    \includegraphics[width=\linewidth]{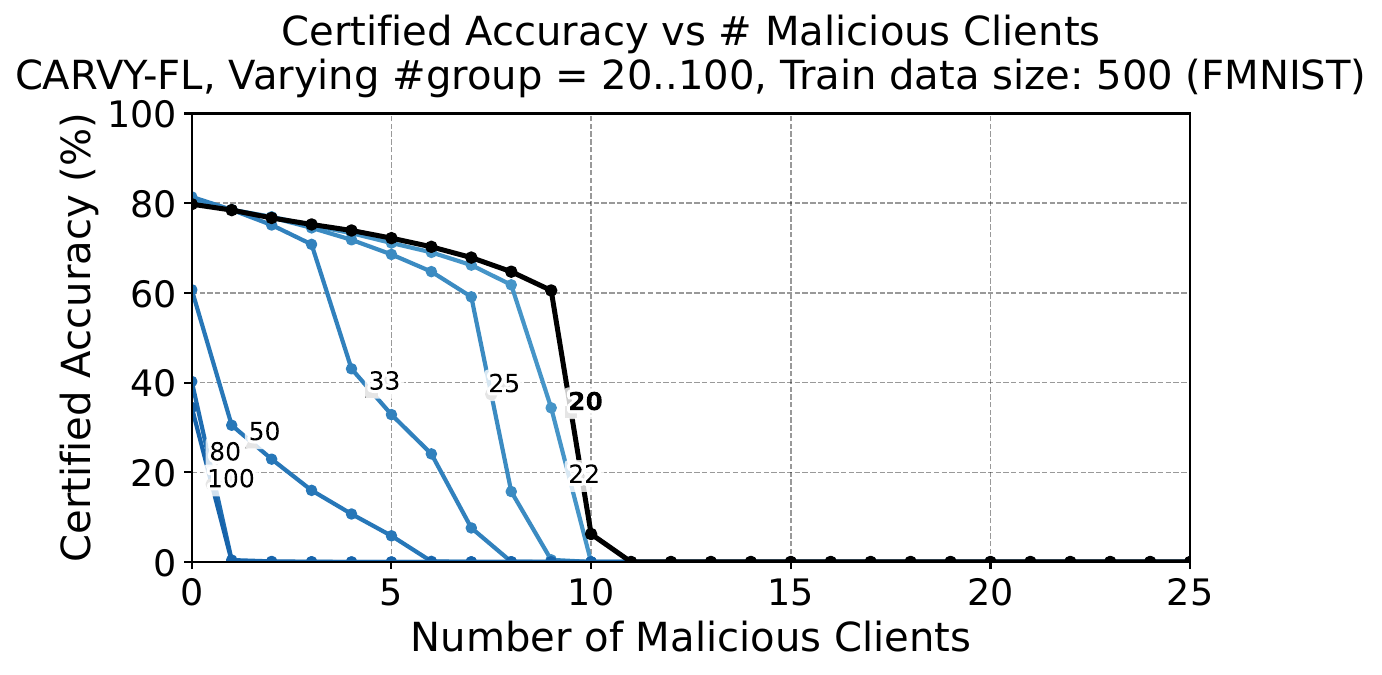}
    \caption{FMNIST: $20 \le G \le 100$}
    \label{fig:FMNIST_group_num_CARVY-FL_20over}
  \end{subfigure}

  \caption{Certified accuracy of CARVY-FL under varying numbers of groups (MNIST / FMNIST).}
  \Description{Four line plots showing certified accuracy of CARVY-FL on MNIST and FMNIST for group counts ranging from 1 to 100, split into two ranges at G=20.}
  \label{fig:group_num_CARVY-FL_mnist_fmnist}
\end{figure*}

We present the results and discussion on CA when varying the number of groups.
As shown in Figure~\ref{fig:group_num_CARVY-FL_mnist_fmnist}, the 20-group configuration automatically determined by CARVY-FL yields the highest attack resilience among all evaluated group counts.
Specifically, with 20 groups, the smallest number of malicious clients at which CA drops to zero is maximized, and for all smaller numbers of malicious clients, CA remains equal to or higher than that of other group count configurations.
This trend was consistently observed across both the MNIST and Fashion-MNIST datasets.
Therefore, the group count determination logic of CARVY-FL appropriately selects a group count that achieves high certified accuracy under these experimental conditions.
Furthermore, these results suggest that Anticlustering, the grouping method employed by CARVY-FL, automatically identifies a group count that provides a favorable trade-off between the inference cost proportional to the number of groups and the attack resilience represented by certified accuracy.

\begin{figure*}[t]
  \centering

  \begin{subfigure}[t]{0.49\linewidth}
    \centering
    \includegraphics[width=\linewidth]{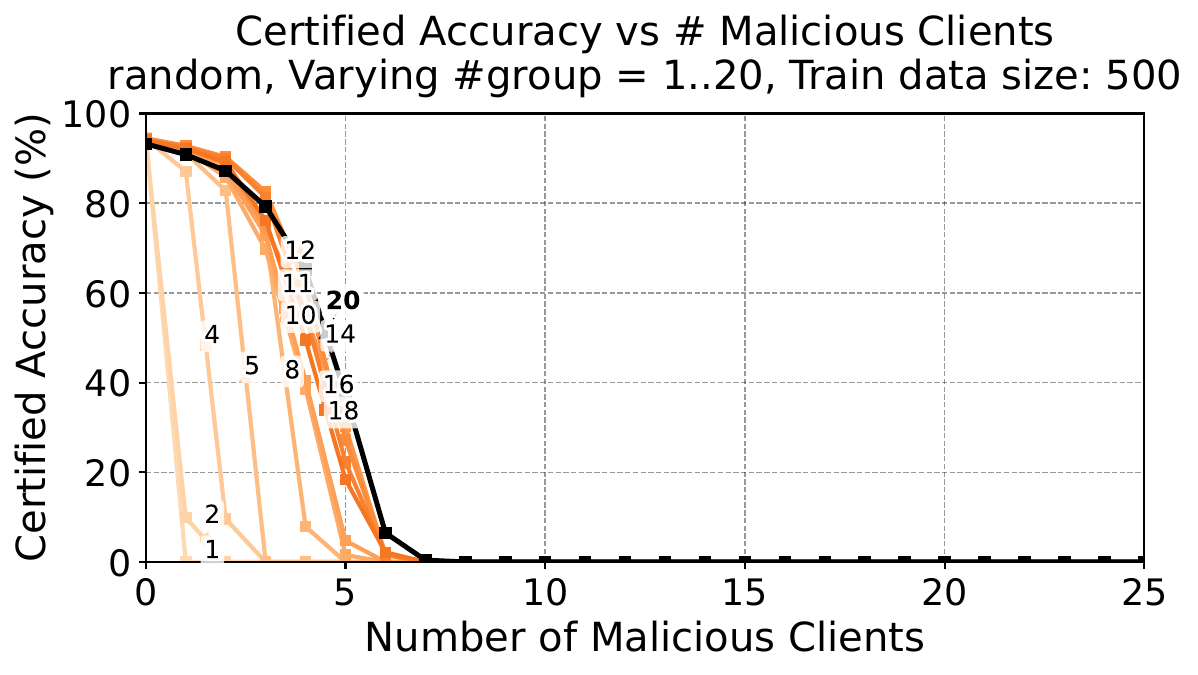}
    \caption{MNIST: $1 \le G \le 20$}
    \label{fig:MNIST_group_num_random_20under}
  \end{subfigure}\hfill
  \begin{subfigure}[t]{0.49\linewidth}
    \centering
    \includegraphics[width=\linewidth]{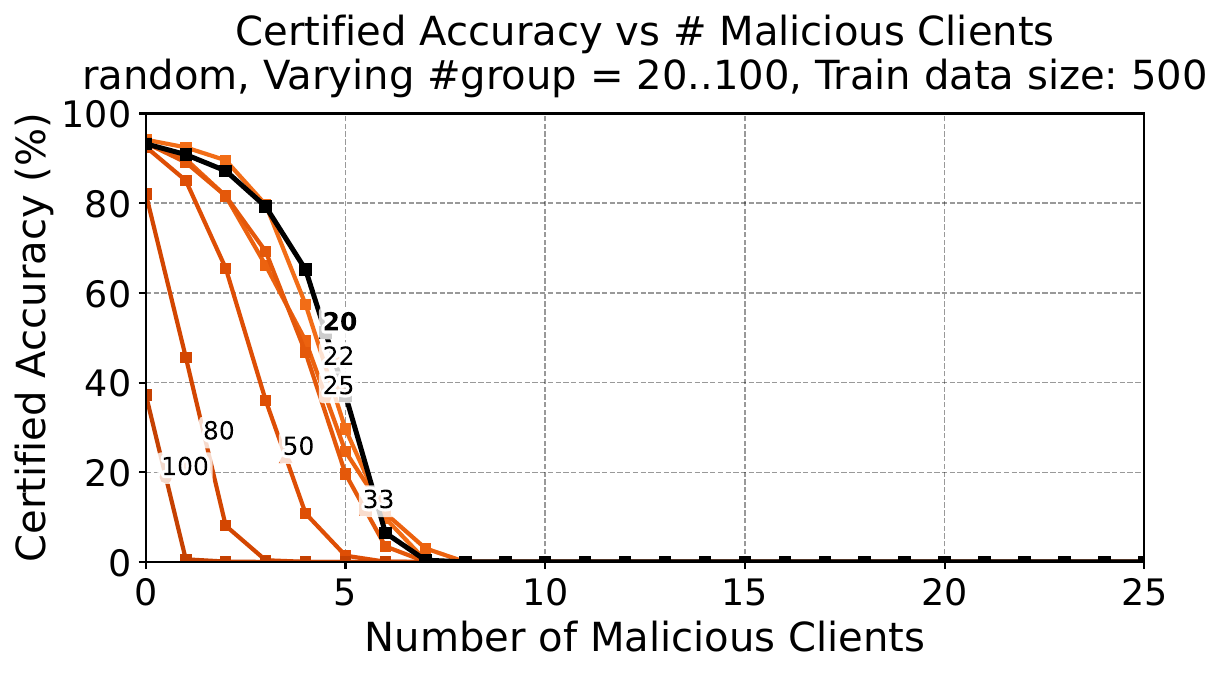}
    \caption{MNIST: $20 \le G \le 100$}
    \label{fig:MNIST_group_num_random_20over}
  \end{subfigure}

  \begin{subfigure}[t]{0.49\linewidth}
    \centering
    \includegraphics[width=\linewidth]{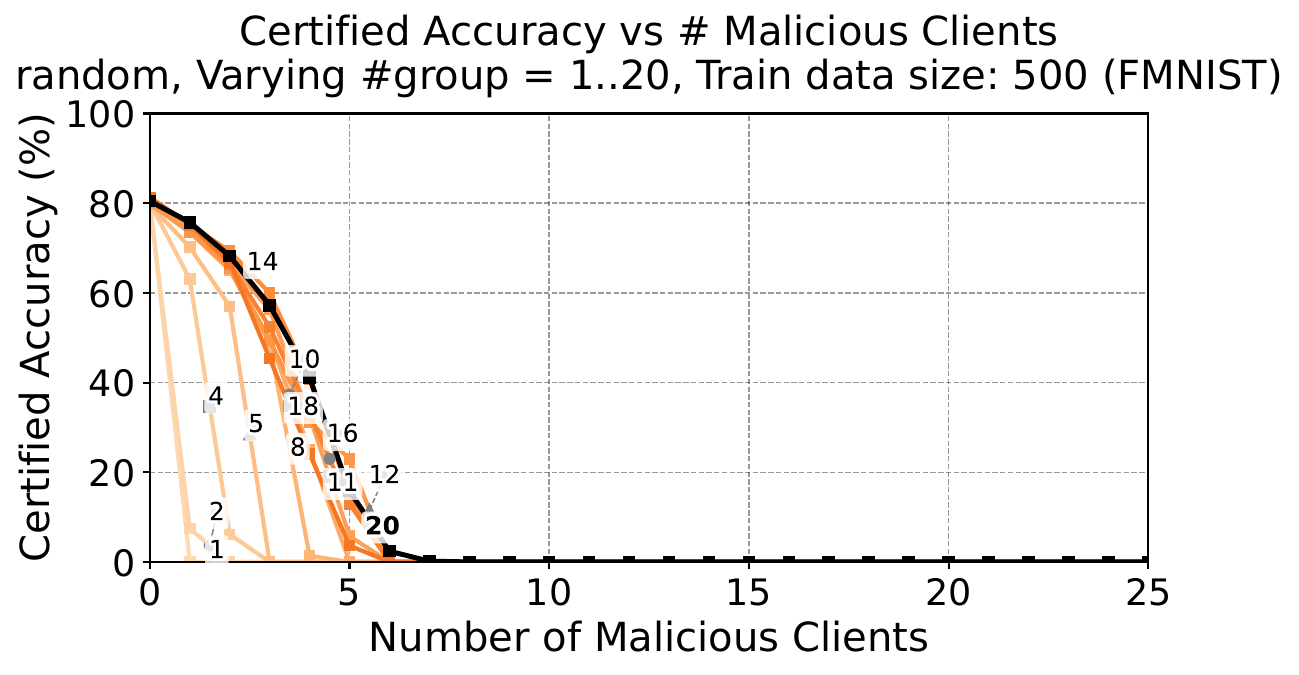}
    \caption{FMNIST: $1 \le G \le 20$}
    \label{fig:FMNIST_group_num_random_20under}
  \end{subfigure}\hfill
  \begin{subfigure}[t]{0.49\linewidth}
    \centering
    \includegraphics[width=\linewidth]{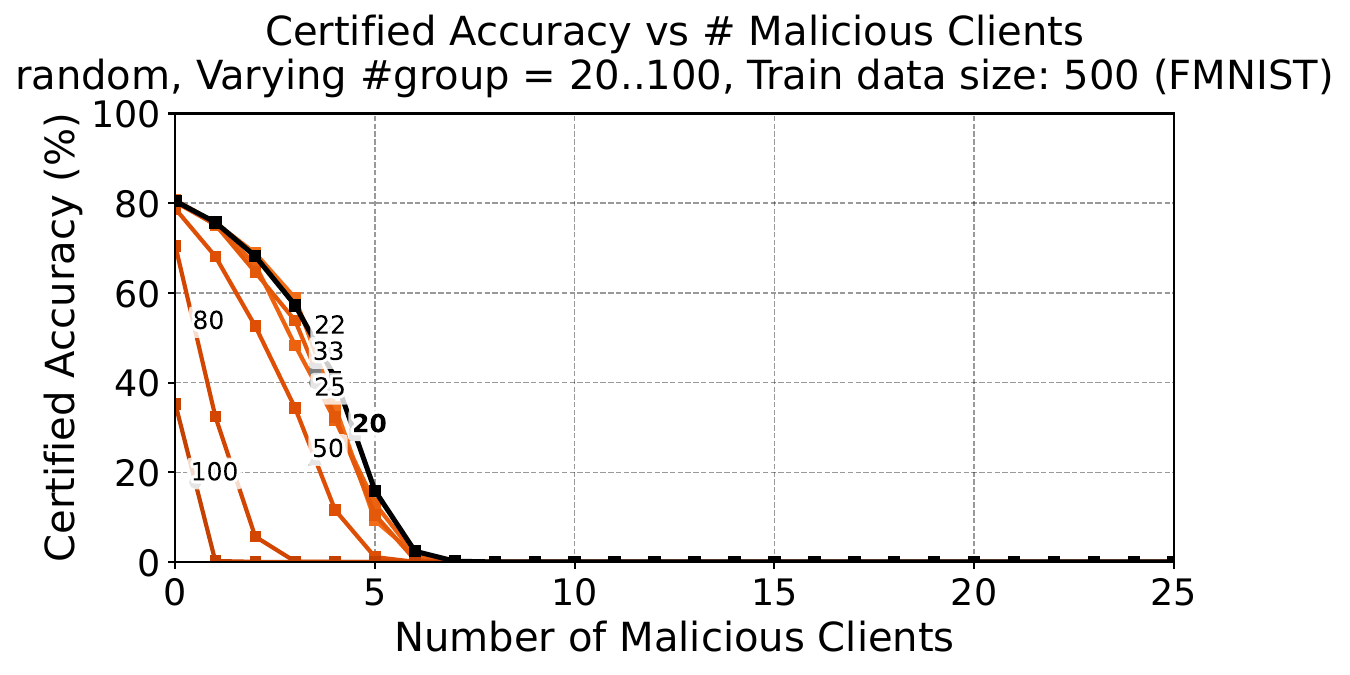}
    \caption{FMNIST: $20 \le G \le 100$}
    \label{fig:FMNIST_group_num_random_20over}
  \end{subfigure}

  \caption{Certified accuracy of FLCert under varying numbers of groups (MNIST / FMNIST).}
  \Description{Four line plots showing certified accuracy of FLCert on MNIST and FMNIST for group counts ranging from 1 to 100, split into two ranges at G=20.}
  \label{fig:group_num_random_mnist_fmnist}
\end{figure*}

Next, Figure~\ref{fig:group_num_random_mnist_fmnist} shows the CA of FLCert under varying numbers of groups.
Although 20 groups does not always yield the best CA at every individual number of malicious clients, it consistently achieves values close to the best.
As shown in Tables~\ref{tab:groupnum_auc_mnist} and \ref{tab:groupnum_auc_fmnist}, the AUC of these plots is maximized at 20 groups.
Therefore, the group count of 20 used for FLCert in the previous subsection is justified.

\begin{figure*}[t]
  \centering

  \begin{subfigure}[t]{0.49\linewidth}
    \centering
    \includegraphics[width=\linewidth]{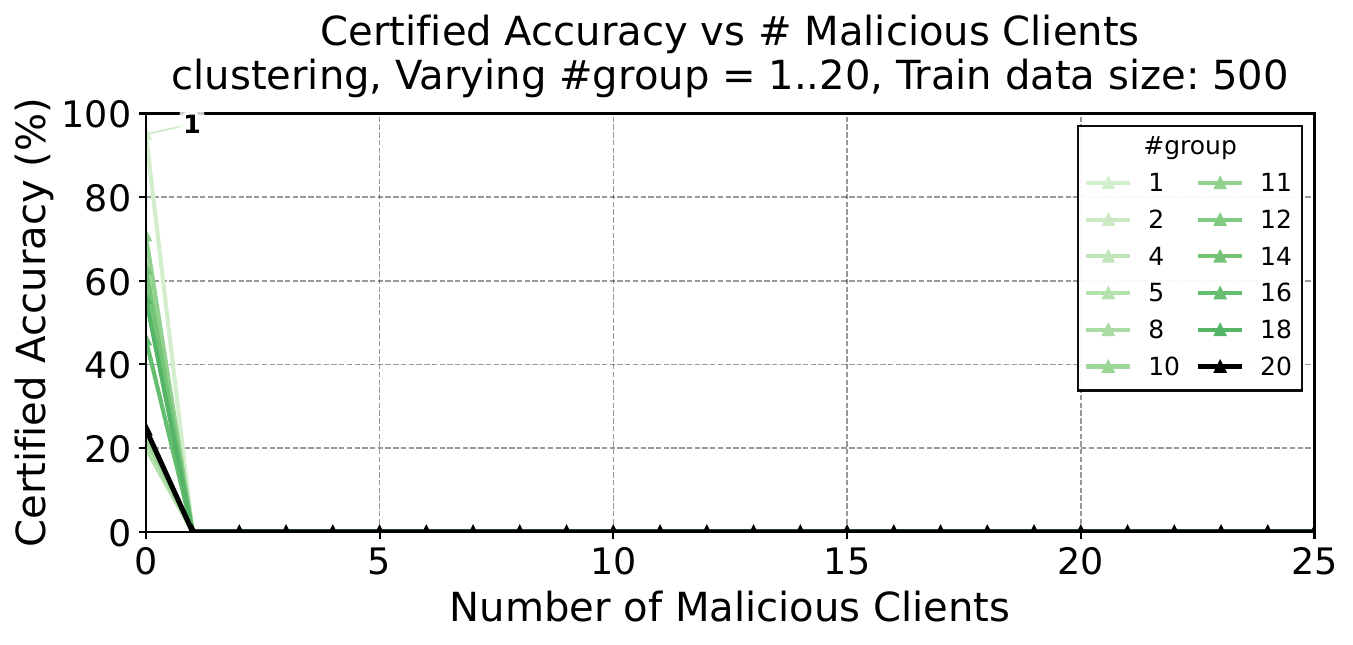}
    \caption{MNIST: $1 \le G \le 20$}
    \label{fig:MNIST_group_num_cluster_20under}
  \end{subfigure}\hfill
  \begin{subfigure}[t]{0.49\linewidth}
    \centering
    \includegraphics[width=\linewidth]{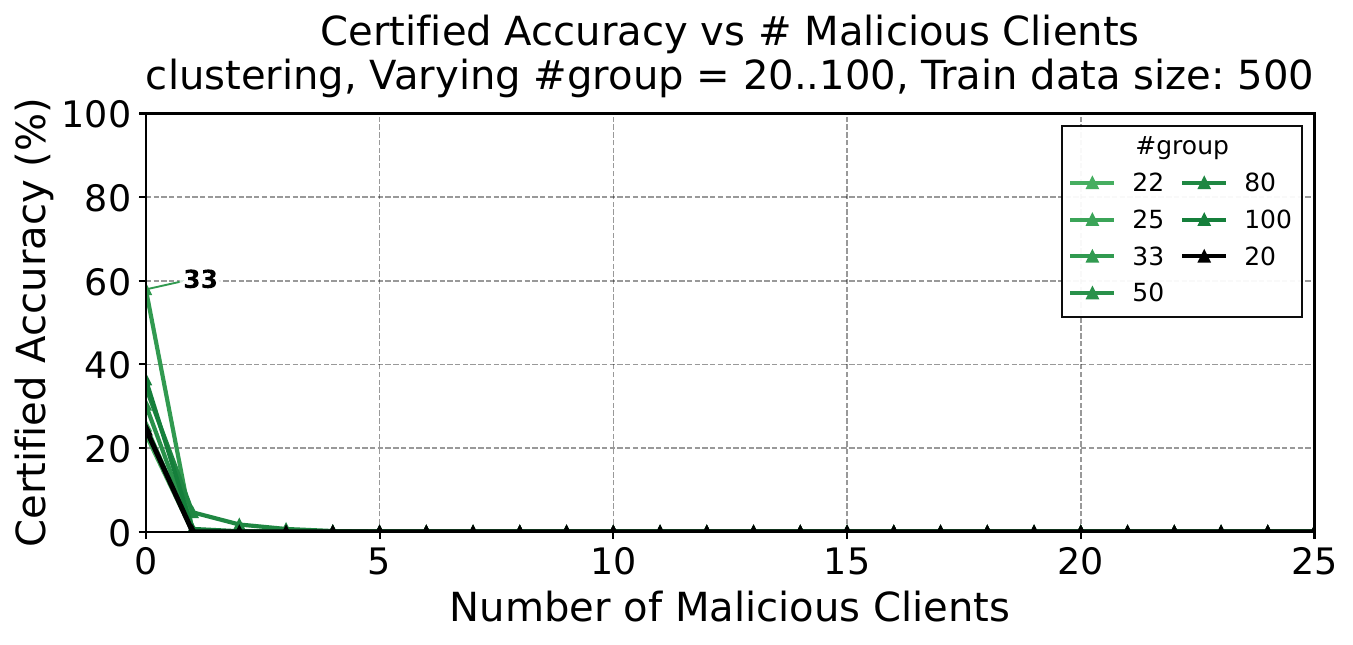}
    \caption{MNIST: $20 \le G \le 100$}
    \label{fig:MNIST_group_num_cluster_20over}
  \end{subfigure}

  \begin{subfigure}[t]{0.49\linewidth}
    \centering
    \includegraphics[width=\linewidth]{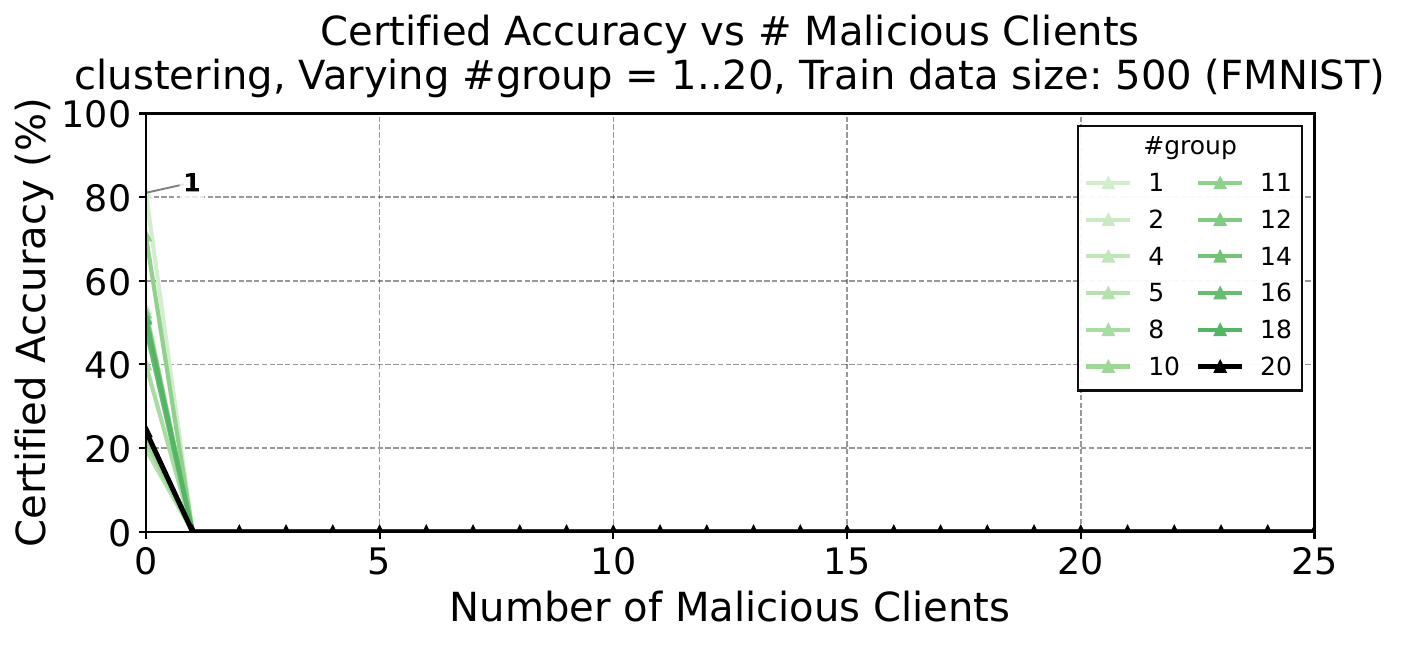}
    \caption{FMNIST: $1 \le G \le 20$}
    \label{fig:FMNIST_group_num_cluster_20under}
  \end{subfigure}\hfill
  \begin{subfigure}[t]{0.49\linewidth}
    \centering
    \includegraphics[width=\linewidth]{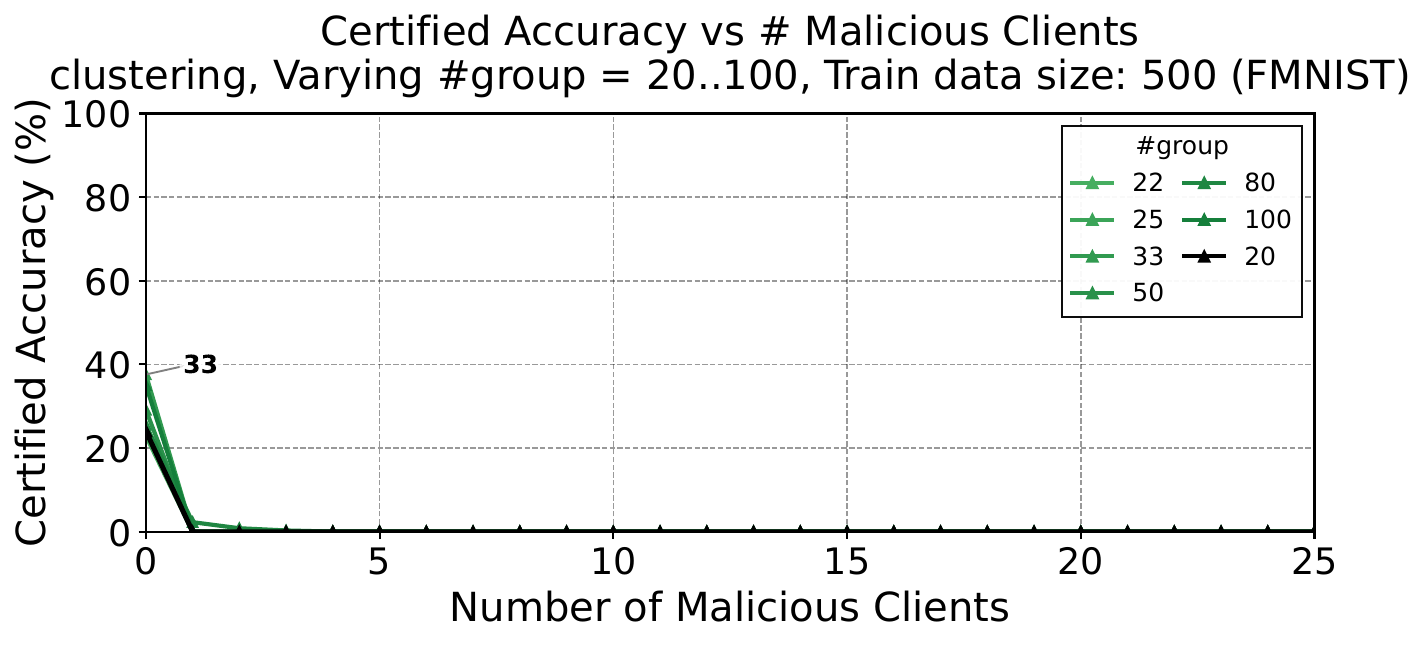}
    \caption{FMNIST: $20 \le G \le 100$}
    \label{fig:FMNIST_group_num_cluster_20over}
  \end{subfigure}

  \caption{Certified accuracy of FLCert+clustering under varying numbers of groups (MNIST / FMNIST).}
  \Description{Four line plots showing certified accuracy of FLCert+clustering on MNIST and FMNIST for group counts ranging from 1 to 100, split into two ranges at G=20.}
  \label{fig:group_num_cluster_mnist_fmnist}
\end{figure*}

Finally, Figure~\ref{fig:group_num_cluster_mnist_fmnist} shows the CA of FLCert+clustering under varying numbers of groups.
The AUC is maximized when the number of groups is 1; however, this configuration corresponds to standard federated learning without FLCert's plurality voting.
Moreover, its AUC is substantially smaller than the maximum AUC achieved by the other two methods at 20 groups, and CA drops to nearly zero whenever even a single malicious client is present, making this configuration impractical for comparison.
Accordingly, the group count was set to 20, consistent with the other two methods, in the previous subsection.

\end{document}